\documentclass[journal]{IEEEtran}
\usepackage{amsmath,amssymb,amsfonts}
\usepackage{algorithm}
\usepackage{algpseudocode}
\makeatletter
\newcommand{\LongState}[1]{\State\parbox[t]{\dimexpr\linewidth-\ALG@thistlm-1em\relax}{#1\strut}}
\makeatother
\usepackage{graphicx}
\usepackage{textcomp}
\usepackage{xcolor}
\usepackage{hyperref}
\hypersetup{hidelinks}
\usepackage{cite}
\usepackage{adjustbox}

\usepackage{makecell}   
\usepackage{arydshln}   
\usepackage{placeins}
\usepackage{braket}     
\usepackage{subcaption}
\usepackage{float}
\usepackage{rotating}   

\newcommand{\mc}[1]{\mathcal{#1}}

\newfloat{protocol}{tbp}{lop}
\floatname{protocol}{Protocol}

\def\BibTeX{{\rm B\kern-.05em{\sc i\kern-.025em b}\kern-.08em
    T\kern-.1667em\lower.7ex\hbox{E}\kern-.125emX}}

\begin{document}

\title{Simulation of Two-Qubit Grover's Algorithm in MBQC with Universal Blind Quantum Computation}

\author{Youngkyung~Lee,
        and~Doyoung~Chung%
\thanks{This work was supported by the Electronics and Telecommunications Research Institute (ETRI) grant funded by the Korean government [26ZS1320, Research on Quantum-Based New Cryptographic System for Ensuring Perfect Data Privacy]. \textit{(Corresponding author: Doyoung Chung.)}}%
\thanks{Youngkyung Lee and Doyoung Chung are with the Cryptography Engineering Laboratory, Electronics and Telecommunications Research Institute (ETRI), Daejeon 34129, South Korea (e-mail: youngklee@etri.re.kr; thisisdoyoung@etri.re.kr).}}

\markboth{}%
{Lee \MakeLowercase{\textit{et al.}}: Simulation of Two-Qubit Grover's Algorithm in MBQC with Universal Blind Quantum Computation}

\IEEEtitleabstractindextext{%
\begin{abstract}
Simulating the Universal Blind Quantum Computation (UBQC) protocol on gate-based platforms requires updating each measurement basis from earlier outcomes within the same execution. Existing Measurement-Based Quantum Computation (MBQC) simulation tools typically handle this feed-forward outside a single executable quantum program, so they do not natively expose the fixed-graph measurement-by-measurement update order studied here. We address this gap by implementing the flow-based Pauli corrections and a simulation-level UBQC angle-update rule as in-circuit dynamic-circuit primitives in Qiskit. Applied to two-qubit Grover's algorithm on a custom $2\times 9$ flow of 18 qubits, the resulting program achieves deterministic success under both plain MBQC and the UBQC layer. Under the depolarizing-noise setting considered here, the angle-blinding layer remains close to plain MBQC, with only a modest additional gap relative to the dominant MBQC-over-circuit overhead. To show that the construction is not tied only to this Grover-specific pattern, we also verify a $G_{2,5}$ brickwork-state fragment with representative universal-gate primitives under the same client/server blinding view. Structural blindness for a full algorithmic brickwork resource remains future work, but within this scope the construction provides a concrete dynamic-circuit route toward fixed-graph, privacy-preserving delegated-quantum-computing simulation at the protocol-update level.
\end{abstract}

\begin{IEEEkeywords}
Grover's algorithm, Measurement-based quantum computation (MBQC), Qiskit, quantum cloud computing, quantum simulation, universal blind quantum computation (UBQC).
\end{IEEEkeywords}}

\maketitle

\IEEEdisplaynontitleabstractindextext
\IEEEpeerreviewmaketitle

\section{Introduction}
\label{sec:intro}

Quantum algorithms such as Shor's factoring~\cite{Shor1997} and Grover's search~\cite{Grover1996, Bennett1997, Boyer1998} have established the promise of quantum computation, and the growing availability of cloud-accessible quantum processors has made delegated quantum computing a practical concern: a delegating client must trust the remote server with its input data, its algorithm, and its output~\cite{Childs2005}. Universal Blind Quantum Computation (UBQC), introduced by Broadbent, Fitzsimons, and Kashefi~\cite{Broadbent2009}, addresses all three by allowing a nearly-classical client to delegate arbitrary computations while revealing none of them to the server. UBQC is built on the Measurement-Based Quantum Computation (MBQC) model~\cite{Raussendorf2001, Briegel2009}, in which computation proceeds through adaptive single-qubit measurements on an entangled resource state, and it has become a foundational framework for privacy-preserving quantum computing~\cite{morimae2013blind, Barz2013, dunjko2014composable, Fitzsimons2017, takeuchi2018verification, Kashefi2016multiparty, Dunjko2012, drmota2023verifiable, Wei2025distributed, Kim2024etri}.

Despite the theoretical equivalence between MBQC and the circuit model~\cite{Raussendorf2003}, simulating MBQC---and, a fortiori, UBQC---on a standard gate-based platform is nontrivial. MBQC is driven by non-unitary measurements whose basis must be adjusted \emph{in real time} from earlier outcomes, following the flow-based correction rule of Danos and Kashefi~\cite{Danos2007}. Reproducing this adaptive feed-forward on a circuit-based simulator therefore requires binding every measurement outcome to the angles that depend on it before the next measurement is issued.

Existing MBQC simulation tools typically keep this outcome-conditional correction \emph{outside} the quantum program. Specialized pattern compilers such as \texttt{graphix}~\cite{graphix}, \texttt{MentPy}~\cite{mentpy}, and the MBQC module of \texttt{Paddle-Quantum}~\cite{paddlemqbc} offer strong graph/flow analysis, but they resolve byproducts as classical post-processing on measurement records or as host-side updates between shots. A gate-level Qiskit demonstration by Kashif and Al-Kuwari~\cite{Kashif2022} likewise stops at cluster-state preparation and static measurement patterns without integrating the in-circuit correction loop (Section~\ref{sec:comparison}). These pipelines can reproduce an MBQC pattern's output distribution, but they do not natively instantiate the same single-program adaptive execution trace as the UBQC-style update order considered here.

This paper studies that missing execution layer in a deliberately restricted setting. For a fixed public graph and a simulation-level client/server encoding, we implement the flow-based Pauli corrections and the BFK09 UBQC angle-update rule as \emph{in-circuit} dynamic-circuit primitives in Qiskit (\texttt{c\_if}). The resulting quantum program follows the BFK09-style measurement-update sequence at simulation level while keeping correction and unblinding inside the same executable circuit. We validate the construction on two-qubit Grover's algorithm~\cite{NielsenChuang} over a custom $2\times 9$ flow of 18 qubits, under both plain MBQC and the BFK09 UBQC layer. Within this setup, the custom flow delivers \emph{fixed-graph angle-based} blindness; \emph{structural} blindness via the universal brickwork state of~\cite{Broadbent2009} requires a larger resource and is deferred to future work (Section~\ref{sec:ubqc_security}).

The main contributions of this paper are as follows:
\begin{itemize}
    \item \textbf{In-circuit correction engine.} We factor the Danos--Kashefi flow-based Pauli correction rule (Eq.~\eqref{eq:correction_sets}) into an in-circuit Qiskit subroutine (Algorithm~\ref{alg:Measurement}) in which XOR-parity corrections are realized as sequences of single-qubit \texttt{c\_if}-conditional Paulis, without runtime classical arithmetic on the output register. We verify the construction end-to-end on the universal gate set $\{H, X, Z, T, CZ\}$ with 1024-shot statistics (Section~\ref{sec:engine}, Appendix~\ref{app:gate_verification}).
    \item \textbf{Fixed-graph $2\times 9$ Grover/UBQC realization.} We construct a flow-compatible 18-qubit $2\times 9$ graph with $VCZ=\{4,5,14,15\}$ and $f(i)=i+2$, so that the oracle and diffusion stages of two-qubit Grover are encoded in local measurement angles on a single resource. This yields deterministic plain-MBQC Grover at the circuit-model $4096/4096$ noise-free success rate, after which the client-side pads of~\cite{Broadbent2009} are added as in-circuit $R_Z(\theta_i)$ preparations and post-measurement $X$-flip re-measurements to obtain a fixed-graph UBQC layer that reproduces the BFK09-style update order and supports fixed-graph angle-based hiding analytically and via output-stage pad suppression (Sections~\ref{sec:grover}--\ref{sec:ubqc}, Appendix~\ref{app:blind_itsec}).
    \item \textbf{Brickwork-fragment portability check.} We verify that the same correction and blinding machinery is not tied only to the Grover-specific flow by simulating a $G_{2,5}$ brickwork-state fragment with representative identity, single-qubit Clifford, entangling, and non-Clifford primitives. The client-decrypted view matches the target logical output, while the server raw view remains uniform under the output one-time pad (Section~\ref{sec:brickwork_fragment}).
    \item \textbf{Noise and scaling analysis.} Under the depolarizing-noise setting studied here, we observe only a modest additional gap from the angle-blinding layer (Fig.~\ref{fig:noise_comparison}), indicating that the UBQC pad operations of Section~\ref{sec:ubqc} contribute less to the total degradation than the underlying MBQC resource overhead in our near-term-style noise model. We also make explicit the $O(nd)$ spacetime scaling of MBQC-on-circuits as a depth--width trade-off for future instantiations (Section~\ref{sec:discussion}).
\end{itemize}

The remainder of this paper is organized as follows. Section~\ref{sec:background} reviews the relevant background on MBQC, BQC, and the Qiskit dynamic-circuit primitives used throughout. Section~\ref{sec:engine} develops the Qiskit-native correction engine and verifies it on the universal gate set. Section~\ref{sec:grover} specializes the engine to the two-qubit Grover's algorithm on the $2\times 9$ flow. Section~\ref{sec:ubqc} layers the BFK09 UBQC protocol on the same circuit, adds the $G_{2,5}$ brickwork-fragment check, and analyzes the resulting blindness. Section~\ref{sec:discussion} reports the noise-robustness, resource-overhead, and tool-comparison results, and Section~\ref{sec:conclusion} concludes. The appendices collect the Pauli-frame derivations, per-gate Qiskit verification figures, full plain and blind Grover circuits, brickwork-fragment angle patterns, and the information-theoretic leakage bound.

\section{Background}
\label{sec:background}

\subsection{Notation}

We adopt standard notation for the single-qubit Pauli operators $X, Z$ and the controlled-$Z$ gate~$CZ$; $R_\sigma(\theta)_i = e^{-i\theta\sigma/2}$ denotes the rotation on qubit $i$ about axis $\sigma \in \{X, Y, Z\}$. Below we collect the paper-specific notation used in Sections~\ref{sec:engine}--\ref{sec:ubqc}.

\begin{itemize}
    \item $[n, m] = \{k \in \mathbb{Z} \mid n \leq k \leq m\}$ denotes an integer interval.

    \item The \emph{tilted basis} is $\{\ket{+_\theta}, \ket{-_\theta}\}$ with
    \[
        \ket{\pm_\theta} = \tfrac{1}{\sqrt{2}}(\ket{0} \pm e^{i\theta}\ket{1});
    \]
    note $\ket{+_0} = \ket{+}$ and $\ket{+_\pi} = \ket{-}$.

    \item $M_i^Z$ denotes a $Z$-basis measurement on qubit $i$; $M_i^\theta$ denotes a measurement in the $\ket{\pm_\theta}$-basis (equivalently, along the axis rotated by $\theta$ from the $X$-axis in the $XY$-plane). The outcome $s_i \in \{0,1\}$ labels $\ket{+_\theta}$ and $\ket{-_\theta}$ respectively.

    \item Byproduct-operator exponents follow the convention $X^s = X$ if $s = 1$ and $X^s = I$ if $s = 0$ (likewise for $Z^s$); sums $\sum_j s_j$ appearing in Pauli exponents are always understood mod~2.

    \item Correction sets $S_X(i), S_Z(i) \subseteq V(G)$ collect the previously measured vertices whose outcomes contribute $X$- and $Z$-byproducts at qubit~$i$; the resulting flow-corrected adaptive measurement angle is $\phi'_i$, defined formally in Eq.~\eqref{phi2} of Section~\ref{sec:flow}.

    \item For readability, normalization constants are occasionally omitted in state expressions.
\end{itemize}

\subsection{MBQC: Measurement-Based Quantum Computation}

In MBQC, computation is driven by single-qubit measurements on an entangled resource state, with each measurement basis determined adaptively from the outcomes of prior measurements~\cite{Raussendorf2003, Browne2005}.

Concretely, an MBQC pattern operates on a \emph{graph state} $\ket{G}$ associated with an open graph $(G, I, O)$ with vertex set $V(G)$ and designated input/output subsets $I, O \subseteq V(G)$: every vertex is prepared in $\ket{+}$ and a $CZ$ gate is applied along every edge, after which the vertices in $V(G)\setminus O$ are measured in tilted bases $\ket{\pm_\phi}$ at algorithm-specified angles $\phi \in [0, 2\pi)$~\cite{Raussendorf2001, Briegel2009}. Because each measurement outcome $s_i$ leaves a residual $X$- and/or $Z$-\emph{byproduct operator} acting on the un-measured qubits, the measurement angle at each subsequent site must be adjusted to absorb these byproducts; the admissible propagation order and the resulting correction rule are captured by the \emph{flow} formalism of Danos and Kashefi~\cite{Danos2007}, which this paper implements explicitly in Section~\ref{sec:flow}.

Once all non-output measurements are completed, the graph state is consumed---hence the term \textit{one-way quantum computation}~\cite{Raussendorf2001, Jozsa2005}; because the full state must be held in memory before consumption, simulating this pattern on a circuit-based platform incurs an $O(nd)$ spacetime overhead, quantified in Section~\ref{sec:resource_overhead}.

\subsection{Blind Quantum Computation}

Blind Quantum Computation (BQC) is a cryptographic protocol that enables a client with limited quantum computational capability to delegate computations to a more powerful quantum server while preserving the privacy of the client's data and algorithm~\cite{Childs2005, Broadbent2009, Fitzsimons2017}. The concept was first introduced by Childs in 2005 with a protocol ensuring \emph{data} privacy but not function privacy~\cite{Childs2005}. In 2009, Broadbent, Fitzsimons, and Kashefi (BFK09) developed the \textit{Universal Blind Quantum Computation} (UBQC) protocol~\cite{Broadbent2009}, building on MBQC to protect both data and function privacy: the client delegates universal computations to a remote server (Bob) without revealing her input, computation details, or output. Subsequent research has extended UBQC toward verifiable variants (VBQC)~\cite{Fitzsimons2017, Barz2013, Kashefi2017, drmota2023verifiable} and demonstrated its feasibility in small-scale experiments~\cite{Barz2012, Wei2025distributed}.

At the core of BFK09 are two independent per-vertex one-time pads. The \emph{basis pad} $\theta \in \{k\pi/4\}_{k=0}^{7}$ rotates each qubit's preparation by $R_Z(\theta)$ and hides the logical measurement angle, while the \emph{outcome pad} $r \in \{0,1\}$ shifts the measurement angle by $\pi r$ and one-time-pads the reported bit. These pads are applied on top of a universal \emph{brickwork state}~$\mc{G}_{n,m}$, whose algorithm-independent topology additionally hides the algorithm identity itself. We accordingly distinguish two sub-regimes of function privacy, following Section~\ref{sec:ubqc_security}: \emph{angle-based blindness}, delivered by the two pads alone and concealing the measurement angles and reported outcomes, and \emph{structural blindness}, delivered by the brickwork layout and additionally concealing the algorithm identity. Our implementation in Section~\ref{sec:ubqc} realizes the former on a custom $2\times 9$ resource; the latter is deferred as future work due to current simulator capacity (Section~\ref{sec:resource_overhead}).

\subsection{Qiskit}

Qiskit is IBM's open-source, circuit-based quantum computing framework~\cite{javadiabhari2024, QiskitGithub, QiskitNature2021} supporting both hardware execution and simulation. The enabling capability for this work is Qiskit's \emph{adaptive programming}: classical registers can receive mid-circuit measurement outcomes, and conditional gates---historically via \texttt{c\_if}, superseded by \texttt{if\_test} in Qiskit~1.0 and later---can apply operations based on those outcomes in real time~\cite{QiskitTextbook, QiskitGithub, Baumer2024qft}. These primitives furnish the runtime substrate required by MBQC's outcome-dependent basis schedule; Section~\ref{sec:meas_sim} details how we use them to realize flow-based byproduct corrections in-circuit.

\section{Engineering a Qiskit-Native MBQC Correction Engine}
\label{sec:engine}

This section develops the simulation machinery used throughout the remainder of the paper. We translate the flow-based correction formula into a \texttt{c\_if}-based dynamic-circuit construction that runs on a standard gate-based backend, and we specialize the construction to every gate in the universal set $\{H,X,Z,T,CZ\}$, reporting 1024-shot verification for each. The resulting resource graphs, measurement angles, and byproduct operators are collected in a single reference table (Table~\ref{tab:pattern_summary}) that is reused verbatim by the 2-qubit Grover simulation in Section~\ref{sec:grover} and by its blind variant in Section~\ref{sec:ubqc}. Standard gate-teleportation and flow-correction results \cite{Raussendorf2001, Danos2007, Cepaite2017} are recalled only to fix notation; the corresponding Pauli-frame derivations are collected in Appendix~\ref{app:pauli_frame}.

Section~\ref{sec:gate_telep} fixes the gate-teleportation identity, Section~\ref{sec:flow} states the flow-based correction rule, Section~\ref{sec:meas_sim} realizes the $X(\phi)$-basis measurement in Qiskit, and Section~\ref{sec:universality} specializes the construction to each universal gate and reports verification data.

\subsection{Gate Teleportation}
\label{sec:gate_telep}

Gate teleportation is the foundational mechanism of MBQC \cite{Raussendorf2001, NielsenChuang, Wallden2018}. Consider an input qubit in an arbitrary state $\ket{\psi}_1$ and an ancilla prepared in $\ket{+}_2$, entangled by a Controlled-$Z$ ($CZ$) gate. Expanding the entangled state $\ket{\psi'}_{12} = CZ(\ket{\psi}_1 \otimes \ket{+}_2)$ in the $\ket{\pm_\theta}$ basis on qubit~1 gives, for any angle $\theta \in [0, 2\pi)$,
\begin{equation}
\begin{aligned}
\ket{\psi'}_{12}
&= \tfrac{1}{\sqrt{2}}\ket{+_\theta}_1 \otimes J(-\theta)_2 \ket{\psi}_2 \\
&\quad + \tfrac{1}{\sqrt{2}}\ket{-_\theta}_1 \otimes X_2 J(-\theta)_2 \ket{\psi}_2,
\end{aligned}
\end{equation}
where $J(\theta) = H R_Z(\theta)$. The left-hand side is independent of $\theta$; the right-hand side only re-expresses the same state in a $\theta$-rotated measurement basis on qubit~1, which becomes a definite branch upon measurement. Thus a measurement of qubit~1 in the $\ket{\pm_\theta}$ basis teleports $J(-\theta)\ket{\psi}$ onto qubit~2, up to a measurement-dependent Pauli $X$ byproduct that must be corrected by a classically conditioned operation.

Any single-qubit unitary $U$ admits a decomposition $U = J(0)\,J(\theta_1)\,J(\theta_2)\,J(\theta_3)$ \cite{Danos2005, Wallden2018}, so the universal gate set $\{H, X, Z, T, CZ\}$ can be realized with constant-depth measurement patterns (at most four measurements per single-qubit gate and six qubits per $CZ$).

\subsection{Flow and Correction in MBQC}
\label{sec:flow}

The flow is a key concept that defines the measurement and correction order in MBQC. We start with an open graph state \((G, I, O)\), where \(G\) is an undirected graph, and \(I\) and \(O\) are subsets of nodes representing inputs and outputs, with \(I^c\) and \(O^c\) as their complements. A flow exists if there is a map \(f : O^c \to I^c\) and a partial order \(\preceq\) over the qubits \cite{Cepaite2017},\cite{Danos2007}.

\begin{itemize}
    \item $x \sim f(x)$ : $x$ and $f(x)$ are neighbors on the graph
    \item $x \preceq f(x)$ : ($f(x)$ is to the future of $x$)
    \item For all $y \sim f(x)$, we have $x \preceq y$ : (any other neighbors of $f(x)$ are to the future of $x$)
\end{itemize}

The partial order $\preceq$ fixes the measurement schedule: each qubit must be measured before any qubit that lies in its forward cone, ensuring that all byproduct dependencies $S_X(i), S_Z(i)$ used to correct qubit $i$'s measurement basis are already known classical outcomes at the moment of its measurement.

\vspace{0.3cm}

Not every open graph admits a flow; however, Danos and Kashefi \cite{Danos2007} established a polynomial-time algorithm that decides the existence of a flow and constructs one when it exists. The $2\times 9$ grid graph used in this work (Section~\ref{sec:grover}) admits a natural left-to-right flow $f(i) = i+2$, as verified in Section~\ref{sec:phi_prime}.

The Pauli $X$ and $Z$ byproducts modify the measurement basis according to

\begin{equation}\label{eq:Mphi_X}
M_i^{\phi_i} X = M_i^{-\phi_i}
\end{equation}

\begin{equation}\label{eq:Mphi_Z}
M_i^{\phi_i} Z = M_i^{\phi_i + \pi}
\end{equation}

In the MBQC framework utilizing a flow function $f$, the byproduct operators accumulated up to qubit $i$ depend on the measurement outcomes of its predecessors. Specifically, the accumulated error on qubit $i$ can be expressed as $X^{s_X} Z^{s_Z}$, where $s_X, s_Z \in \{0,1\}$ represent the parity of measurement outcomes contributing to bit-flip and phase-flip errors, respectively.

From Eqs.~\eqref{eq:Mphi_X} and \eqref{eq:Mphi_Z}, a measurement at angle $\phi_i$ preceded by a Pauli byproduct $X^{s_X} Z^{s_Z}$ can be rewritten as a measurement at a \textit{shifted} angle acting on the corrected state. Applying Eq.~\eqref{eq:Mphi_X} first followed by Eq.~\eqref{eq:Mphi_Z}, we obtain
\begin{equation}\label{eq:correction_derive}
\begin{aligned}
M_i^{\phi_i}\, X^{s_X} Z^{s_Z}
&= M_i^{(-1)^{s_X}\phi_i}\, Z^{s_Z} \\
&= M_i^{(-1)^{s_X}\phi_i \,+\, \pi s_Z}.
\end{aligned}
\end{equation}
Thus the \textit{effective} measurement angle that implements the intended logical rotation $\phi_i$ is $\phi'_i = (-1)^{s_X}\phi_i + \pi s_Z$. The flow structure supplies the concrete dependencies: a prior measurement outcome $s_j$ contributes an $X$-byproduct to the flow successor $f(j)$ and a $Z$-byproduct to every other graph-neighbor of $f(j)$, where $N_G(v)$ denotes the open neighborhood of vertex $v$ in $G$. Therefore the accumulated byproducts at the receiver qubit $i$ are obtained by summing over all prior vertices whose propagated byproducts land on $i$:

\begin{equation}\label{phi1}
\phi'_i
= (-1)^{\sum_{j:\, f(j)=i} s_j}\phi_i
  + \pi\!\left(\sum_{j:\, i\in N_G(f(j))\setminus\{j\}} s_j\right),
\end{equation}

where \( s_j \) represents the outcome of the measurement on qubit \( j \) \cite{Danos2005, Wallden2018}. The first sum collects the $X$-dependencies of qubit~$i$, while the second sum collects the $Z$-dependencies induced by neighboring flow images.

Alternatively, this correction can be expressed more generally by defining \( S_X(i) \) and \( S_Z(i) \) as the explicit dependency sets for qubit \( i \):

\begin{equation}\label{phi2}
\phi'_i = (-1)^{\sum_{j \in S_X(i)} s_j} \phi_i + \pi \left( \sum_{j \in S_Z(i)} s_j \right).
\end{equation}

Here, $\sum_{j \in S_X(i)} s_j$ and $\sum_{j \in S_Z(i)} s_j$ correspond to the values $s_X$ and $s_Z$ described above.

\vspace{0.3cm}

Implementing these corrections during MBQC simulation requires incorporating the outcomes of prior measurements adaptively; Section~\ref{sec:meas_sim} introduces the concrete Qiskit mechanism used throughout this work.

\subsection{MBQC Measurement Simulation using Qiskit}
\label{sec:meas_sim}

MBQC requires both $Z$-basis measurements $M_i^Z$ in the $\{\ket{0}, \ket{1}\}$ basis and $X(\theta)$-basis measurements in the $\{\ket{+_\theta}, \ket{-_\theta}\}$ basis, where the latter is obtained by rotating the $X$-axis by $\theta$ around the $Z$-axis in the $XY$-plane \cite{Raussendorf2001}. Qiskit and other circuit-based platforms support only $Z$-basis measurements; we therefore simulate an $X(\phi_i)$-basis measurement by pre-rotating the target qubit with $R_Z(-\phi_i)$ followed by $H$ and then performing a $Z$-basis measurement:

\begin{equation}\label{MphiMz}
M_i^{\phi_i} = M_i^Z H R_Z(-\phi_i).
\end{equation}

For example, an $X$-measurement corresponds to \( M_i^Z H \), an \( X(\pi) \)-measurement corresponds to \( M_i^Z H Z \), and an \( X(\pi/4) \)-measurement corresponds to \( M_i^Z H R_Z(-\pi/4) \).

In MBQC simulations in Qiskit, referring to Eqs.~\eqref{phi2} and \eqref{MphiMz}, measurement angles modified by corrections (denoted as \( \phi' \)) can be determined as follows:

\begin{equation}
\begin{aligned}
M_i^{\phi_i'} &= M_i^{\phi_i} X^{\sum_{j \in S_X(i)} s_j} Z^{\sum_{j \in S_Z(i)} s_j}\label{Mphi} \\
&= M_i^Z H R_Z (-\phi_i) X^{\sum_{j \in S_X(i)} s_j} Z^{\sum_{j \in S_Z(i)} s_j},
\end{aligned}
\end{equation}

This expression shows that arbitrary measurements \( M_i^{\phi'_i} \) can be simulated in Qiskit using available gates. Specifically, to measure qubit \( i \), conditional gate applications based on prior measurement outcomes (\( \sum_{j \in S_X(i)} s_j \) and \( \sum_{j \in S_Z(i)} s_j \)) are required.
This can be implemented using Qiskit's adaptive programming functions, in particular \texttt{circuit.<gate>(i).c\_if(c, 1)}, which applies \texttt{<gate>} on qubit $i$ conditioned on the classical register $c$ holding the value $1$. The XOR-parity sums $\sum_{j} s_j \pmod{2}$ appearing in Eq.~\eqref{phi2} are realized by issuing one \texttt{c\_if}-conditioned $X$ (resp.~$Z$) gate per dependency $j \in S_X(i)$ (resp.~$S_Z(i)$); since $X^2 = Z^2 = I$, the composed effect of these single-bit conditionals on the target qubit equals $X^{\bigoplus_{j\in S_X(i)} s_j}$ (resp.~$Z^{\bigoplus_{j\in S_Z(i)} s_j}$), so no runtime classical arithmetic on the register is required. Algorithm~\ref{alg:Measurement} in Section~\ref{sec:grover} formalizes this conditional-correction procedure, which is reused throughout the Grover simulation in Section~\ref{sec:grover} and the blind variant in Section~\ref{sec:ubqc}.

\subsection{Simulating Universality of MBQC with Qiskit}
\label{sec:universality}

Each gate in the universal set $\{H, X, Z, T, CZ\}$ is realized as a constant-depth MBQC measurement pattern executed on Qiskit. Table~\ref{tab:pattern_summary} records the resource graphs, measurement angles, and byproduct operators; the corresponding Pauli-frame derivations are given in Appendix~\ref{app:pauli_frame} and the Qiskit verification figures (resource graph, dynamic circuit, and 1024-shot outcome distribution for each gate) are collected in Appendix~\ref{app:gate_verification}. For each gate we report below the resource graph with input/output assignment, the measurement schedule, and the headline 1024-shot statistic.\\

\subsubsection{Hadamard Gate}

\textbf{Pattern.} A two-qubit chain $q_0$--$q_1$ with $q_0$ carrying $\ket{\psi}$ and $q_1$ prepared in $\ket{+}$; measure $q_0$ along $X(0)$ and read out $q_1$. The byproduct $X^{s_0}$ is corrected by a classically controlled $X$ on $q_1$. (Derivation: Appendix~\ref{app:pauli_frame}.)

\textbf{Verification.} The Qiskit realization (Fig.~\ref{fig:H_gate} in Appendix~\ref{app:gate_verification}) uses the equivalent dynamic circuit with a \texttt{c\_if}-based $X$ correction; with $q_0$ initialized in $\ket{0}$, $q_1$ collapses to $\ket{0}$ or $\ket{1}$ with empirical frequencies $509/1024 \approx 0.497$ and $515/1024 \approx 0.503$ (binomial $1\sigma \approx 0.016$).

\subsubsection{X Gate}

\textbf{Pattern.} A three-qubit chain $q_0$--$q_1$--$q_2$, with the input state $\ket{\psi}$ loaded on $q_0$ and ancillas $q_1, q_2$ prepared in $\ket{+}$; measurement angles $(\phi_0, \phi_1) = (0, \pi)$; the readout is $q_2$, with byproducts $X_2^{s_1}$ and $Z_2^{s_0}$ (up to a global phase $-1$). (Derivation: Appendix~\ref{app:pauli_frame}.)

\textbf{Verification.} The Qiskit realization (Fig.~\ref{fig:X_gate} in Appendix~\ref{app:gate_verification}) maps the input $\ket{0}$ to $\ket{1}$ deterministically under the listed corrections (observed $1024/1024$ shots).

\subsubsection{\texorpdfstring{\( R_Z(\theta) \)-Gate: \textit{Z} and \textit{T} as Special Cases}{RZ(theta)-Gate: Z and T as Special Cases}}

\textbf{Pattern.} A three-qubit chain $q_0$--$q_1$--$q_2$, with the input state $\ket{\psi}$ loaded on $q_0$ and ancillas $q_1, q_2$ prepared in $\ket{+}$; measurement angles $(\phi_0, \phi_1) = (-\theta, 0)$; the readout is $q_2$, with byproducts $X_2^{s_1}$ and $Z_2^{s_0}$. Specializing $\theta = \pi$ gives the $Z$-gate and $\theta = \pi/4$ gives the $T$-gate. (Derivation: Appendix~\ref{app:pauli_frame}.)

\textbf{Verification.} The Qiskit realizations are reported in Figs.~\ref{fig:Z_gate} and \ref{fig:T_gate} (Appendix~\ref{app:gate_verification}). For the $Z$-gate test ($q_0$ initialized in $\ket{+}$), $q_2$ is observed in $\ket{1}$ under $X$-basis readout with empirical frequency $1024/1024$ (theory: $1.000$). For the $T$-gate test ($q_0$ initialized in $\ket{+}$), $q_2 \to \ket{+_{\pi/4}}$ and is observed in $\ket{0}$ with empirical frequency $873/1024 \approx 0.853$ (theory: $\cos^2(\pi/8) \approx 0.8536$; binomial $1\sigma \approx 0.011$).

\subsubsection{CZ Gate}

\textbf{Pattern.} A $2\times 3$ grid on qubits $q_0,\dots,q_5$ with input on $q_0, q_1$ and readout on $q_4, q_5$ (first row $q_0, q_2, q_4$; second row $q_1, q_3, q_5$; entanglement edges as in Fig.~\ref{fig:CZ_gate_sub1}). Measure $q_0, q_1, q_2, q_3$ along $X(0)$. The byproducts on the readout qubits are $X_4^{s_2}, X_5^{s_3}, Z_4^{s_0 \oplus s_3}, Z_5^{s_1 \oplus s_2}$. (Derivation via commuting Pauli byproducts through $CZ$: Appendix~\ref{app:pauli_frame}.)

\textbf{Verification.} The Qiskit realization (Fig.~\ref{fig:CZ_gate} in Appendix~\ref{app:gate_verification}) uses input $\ket{1}\otimes\ket{+}$ and produces output $\ket{1}\otimes\ket{-}$, yielding a 1024-shot deterministic outcome $\ket{11}$ on $(q_4, q_5)$.

\subsubsection{Summary of MBQC Measurement Patterns}
\label{sec:pattern_summary}

Table~\ref{tab:pattern_summary} consolidates the patterns of this section. Each single-qubit gate uses at most two measurements on a chain of three qubits, and the $CZ$-gate uses four measurements on a $2\times 3$ grid. The $X$- and $Z$-byproduct operators in the last two columns follow from the flow-based correction formula Eq.~\eqref{phi2} specialized to each subgraph.

\begin{table*}[tbp]
\centering
\caption{Summary of the MBQC patterns used for the universal gate set $\{H, X, Z, T, CZ\}$. Here $s_j$ denotes the pre-correction outcome of the $X(\phi_j)$ measurement on $q_j$, and ``output'' denotes the unmeasured qubit(s) after the listed Pauli corrections.}
\label{tab:pattern_summary}
\begin{tabular}{lcccll}
\hline
Gate & Resource graph & \#\,qubits & Measurement angles & $X$-correction on output & $Z$-correction on output \\
\hline
$H$    & $q_0$--$q_1$ (chain)                      & 2 & $\phi_0 = 0$                         & $X^{s_0}$ & $-$ \\
$X$    & $q_0$--$q_1$--$q_2$ (chain)               & 3 & $\phi_0 = 0$, $\phi_1 = \pi$         & $X^{s_1}$ & $Z^{s_0}$ \\
$Z$    & $q_0$--$q_1$--$q_2$ (chain)               & 3 & $\phi_0 = -\pi$, $\phi_1 = 0$        & $X^{s_1}$ & $Z^{s_0}$ \\
$T$    & $q_0$--$q_1$--$q_2$ (chain)               & 3 & $\phi_0 = -\pi/4$, $\phi_1 = 0$      & $X^{s_1}$ & $Z^{s_0}$ \\
$R_Z(\theta)$ & $q_0$--$q_1$--$q_2$ (chain)        & 3 & $\phi_0 = -\theta$, $\phi_1 = 0$     & $X^{s_1}$ & $Z^{s_0}$ \\
$CZ$   & $2\times 3$ grid ($q_0,\dots,q_5$)        & 6 & $\phi_0=\phi_1=\phi_2=\phi_3=0$      & $X_4^{s_2}, X_5^{s_3}$ & $Z_4^{s_0\oplus s_3}, Z_5^{s_1\oplus s_2}$ \\
\hline
\end{tabular}
\end{table*}

\section{Two-Qubit Grover's Algorithm in MBQC using Qiskit}
\label{sec:grover}

We now apply the correction engine of Section~\ref{sec:engine} to a non-trivial benchmark: a deterministic, feed-forward MBQC realization of the two-qubit Grover algorithm on Qiskit. The $2\times 9$ grid resource state consolidates all four logical stages of the algorithm --- oracle $X$ wrap, oracle $CZ$, diffusion Hadamards, and diffusion $Z\text{-}CZ\text{-}Z$ --- into a single graph whose flow function $f(i)=i+2$ directly induces the adaptive measurement schedule. This section fixes the resource graph and qubit indexing (Section~\ref{sec:grover_init}), specifies the logical measurement angles $\phi_i$ for each oracle case (Section~\ref{sec:grover_phi}), derives the adaptive angles $\phi'_i$ from the grid topology (Section~\ref{sec:phi_prime}), and reports the 1024-shot Qiskit results for all four oracles (Section~\ref{sec:grover_impl}).

\subsection{Circuit-Based Two-Qubit Grover's Algorithm}
\label{sec:grover_circuit}

\begin{figure}[tbp]
    \centering
    \includegraphics[width=0.94\columnwidth]{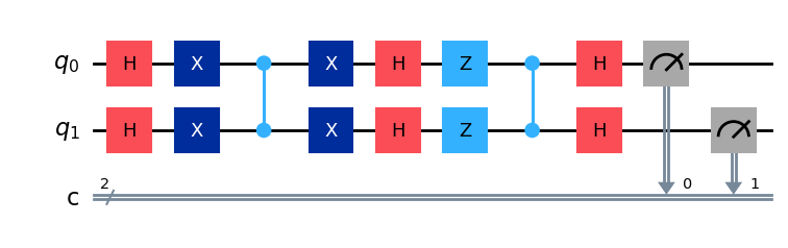}
    \caption{Circuit-model two-qubit Grover algorithm for oracle ``00'': state preparation, oracle, and diffusion. For $n=2$, one Grover iteration suffices and the final $Z$-basis readout on $(q_0,q_1)$ returns the marked bitstring with probability~$1$.}
    \label{fig:2_Qubit_Grover}
\end{figure}

We take as reference the standard two-qubit Grover circuit \cite{Grover1996, NielsenChuang} shown in Fig.~\ref{fig:2_Qubit_Grover}, which organizes into three logical stages: (i)~\emph{state preparation} ($H^{\otimes 2}$ on $\ket{00}$), (ii)~\emph{oracle} (a $CZ$ wrapped by $X$-gates on the qubits whose marked-bitstring value is $0$), and (iii)~\emph{diffusion} ($H^{\otimes 2}$, $Z^{\otimes 2}$, $CZ$, $Z^{\otimes 2}$, $H^{\otimes 2}$). For $n=2$ the algorithm requires only a single Grover iteration and outputs the marked bitstring with probability~$1$, which serves as the deterministic ground truth for the MBQC realization below. The four oracle settings are selected by the $X$-wrap pattern around the oracle $CZ$: the displayed circuit marks ``00''; omitting both wraps marks ``11''; keeping the wrap on only the mismatched qubit marks ``01'' or ``10''.\\

\subsection{Initial State Setup for MBQC Simulation}
\label{sec:grover_init}

\begin{figure}[tbp]
    \centering
        \includegraphics[width=\columnwidth]{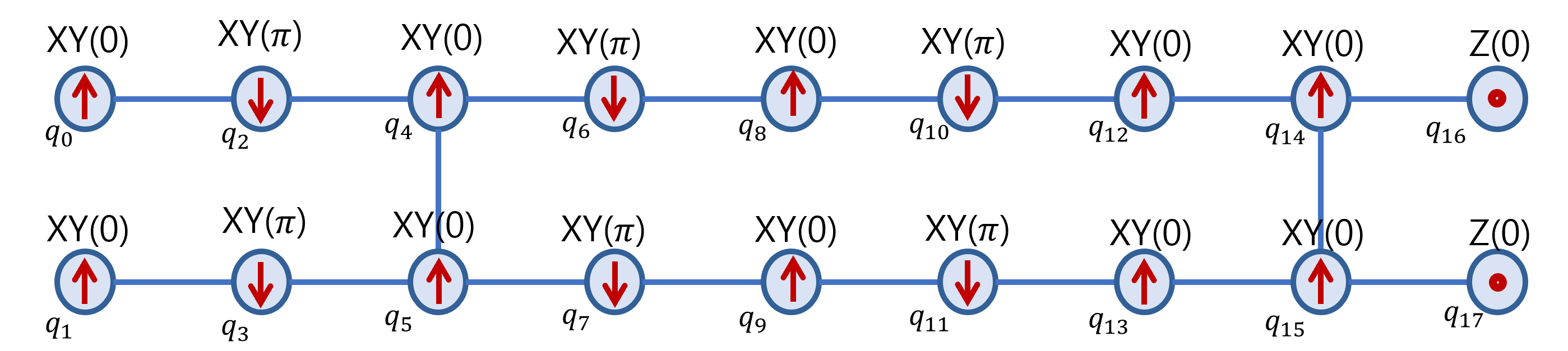}
        \caption{$2\times 9$ MBQC measurement pattern for two-qubit Grover (oracle ``00''). Inputs are $q_0,q_1$, outputs are $q_{16},q_{17}$, horizontal edges carry the row-wise $CZ$ entanglement, and the vertical edges $\{(q_4,q_5),(q_{14},q_{15})\}$ implement the oracle and diffusion entangling gates. Arrows indicate the logical angles $\phi_i \in \{0,\pi\}$; adaptive shifts to $\phi'_i$ are applied at run time via \texttt{c\_if}.}
        \label{fig:MBQC_Grover}
\end{figure}

We realize the two-qubit Grover algorithm on a total of $18$ qubits arranged as a $2\times 9$ grid (Fig.~\ref{fig:MBQC_Grover}). For consistency with Qiskit's flat index convention we use the bijection $q_{x,y} = q_{2y+x}$ with $x \in \{0,1\}$ and $y \in \{0, \dots, 8\}$, so that row $x$ of column $y$ carries the index $2y+x$.

The $2\times 9$ shape is a compact two-row ladder resource satisfying the following three MBQC-level constraints: (i)~the horizontal row-wise flow $f(i)=i+2$ must be preserved end-to-end so that Eq.~\eqref{eq:correction_sets} defines a valid correction schedule on every measured vertex; (ii)~the Grover pattern requires exactly two entangling ``columns'' where a $CZ$ is placed across the two rows---one for the oracle and one for the diffusion operator---so the vertical-edge set $E_{\text{vert}}$ must have cardinality at least two, realised here at $\{(q_4,q_5),(q_{14},q_{15})\}$; and (iii)~the oracle $X$-wraps require two three-site row chains per rail ($q_0$--$q_2$--$q_4$ and $q_4$--$q_6$--$q_8$ on the upper rail, with the analogous odd-indexed chains on the lower rail), while the diffusion block occupies the remaining row segment through the output column. Under these design constraints, reducing the grid width would either break the chosen flow condition or merge the oracle and diffusion stages into a single $CZ$ column in a way that conflicts with the present $X$-wrap construction. Compared with a BFK09 brickwork state $\mc{G}_{n,m}$ of width $m = 4(N_{2q}+k)+1$ (Section~\ref{sec:resource_overhead}), the $2\times 9$ ladder is an algorithm-specific, structurally-transparent resource: it is chosen here \emph{not} as a weaker substitute for a universal brickwork, but as a compact resource on which the in-circuit correction engine of Section~\ref{sec:engine} and the angle-blindness layer of Section~\ref{sec:ubqc} can be exercised within current statevector-simulator capacity. Following the open-graph notation of Section~\ref{sec:flow}, our resource is the triple $(G, I, O)$ with vertex set $V(G) = \{q_0, \dots, q_{17}\}$, input $I = \{q_0, q_1\}$, output $O = \{q_{16}, q_{17}\}$, and edge set
\begin{equation}
E(G) = E_{\text{horiz}} \cup E_{\text{vert}},
\label{eq:grover_edges}
\end{equation}
where $E_{\text{horiz}} = \{(q_{2y+x}, q_{2(y+1)+x}) : x\in\{0,1\}, y\in\{0,\dots,7\}\}$ covers all horizontal adjacencies in both rows, and $E_{\text{vert}} = \{(q_4, q_5), (q_{14}, q_{15})\}$ contains the two vertical edges that realize the entangling gates of the oracle and of the diffusion operator (deferred to Section~\ref{sec:grover_phi}). The resource state is prepared by initializing every qubit in $\ket{+}$ and applying a $CZ$ along each edge of $E(G)$.\\

\subsection{\texorpdfstring{Measurement Angle $\phi_i$}{Measurement Angle phi i}}
\label{sec:grover_phi}

The 16 non-output qubits $q_0, \dots, q_{15}$ are measured sequentially in ascending index order, each with a logical angle $\phi_i \in \{0, \pi\}$; the corresponding adaptive angle $\phi'_i$ derived from the flow is obtained in Section~\ref{sec:phi_prime}. By the gate-teleportation identity $J(\phi) = H R_Z(\phi)$ of Section~\ref{sec:gate_telep}, $\phi = 0$ implements $J(0) = H$ and $\phi = \pi$ implements $J(\pi) = HZ$. Combined with the identity $X = HZH$, two successive measurements along a chain at angles $(0, \pi)$ realize the circuit-model $X$-gate (Section~\ref{sec:universality}, Table~\ref{tab:pattern_summary}); this is the building block through which the oracle's $X$-wrap is mapped onto measurement angles.

All logical angles used here lie in $\{0,\pi\} \subset \{k\pi/4\}_{k=0}^{7}$, the $\pi/4$-grid on which the UBQC blinding of Section~\ref{sec:ubqc} operates; this inclusion is what lets the same pattern be delegated blindly in Section~\ref{sec:ubqc} without enlarging the angle alphabet. Writing $v_1v_0$ for the displayed marked bitstring in the Qiskit output order $q_{17}q_{16}$, let $w_x = 1 - v_x$ indicate whether an $X$-wrap is required on input qubit $q_x$ (so $w_x = 1$ when the corresponding bit of the oracle is $0$). The 16 logical angles then decompose into four functional blocks as summarized in Table~\ref{tab:grover_phi}.

\begin{table}[tbp]
\centering
\caption{Logical measurement angles $\phi_i$ for two-qubit Grover on the $2\times 9$ grid. Here $w_x=1$ iff the oracle requires an $X$-wrap on input qubit $q_x$; the output qubits $q_{16},q_{17}$ are omitted because they are read out in the $Z$ basis after correction.}
\label{tab:grover_phi}
\begin{tabular}{llcc}
\hline
Block & Qubits & $\phi_i$ & Oracle-dependent?\\
\hline
Oracle $X$-wrap (front) & $q_0, q_1$                  & $0$                    & no  \\
Oracle $X$-wrap (front) & $q_2, q_3$                  & $w_0 \pi,\; w_1 \pi$   & \textbf{yes} \\
Oracle $CZ$ edge        & $q_4, q_5$                  & $0$                    & no  \\
Oracle $X$-wrap (back)  & $q_6, q_7$                  & $w_0 \pi,\; w_1 \pi$   & \textbf{yes} \\
Diffusion $H$           & $q_8, q_9$                  & $0$                    & no  \\
Diffusion $Z$           & $q_{10}, q_{11}$            & $\pi$                  & no  \\
Diffusion $CZ$ edge     & $q_{12}, q_{13}$            & $0$                    & no  \\
Diffusion $H$ (final)   & $q_{14}, q_{15}$            & $0$                    & no  \\
\hline
\end{tabular}
\end{table}

Two mappings are worth highlighting. First, the circuit-model initial $H^{\otimes 2}$ is absorbed into the resource-state preparation ($\ket{+}$ on every vertex), so no separate qubits are allocated for it. Second, the oracle's ``$X$-wrap around $CZ$'' decomposes into two $X$-gates (front wrap at $q_2, q_3$ and back wrap at $q_6, q_7$) that sandwich the vertical $CZ$ edge $(q_4, q_5)$; since $X = HZH$, each $X$ is realized as an angle-$\pi$ measurement on the intermediate qubit of a three-qubit chain, which is exactly the pattern of Section~\ref{sec:universality}. Fig.~\ref{fig:MBQC_Grover} visualizes the resulting pattern for oracle $= $ ``00'' (i.e.\ $w_0 = w_1 = 1$); the other three oracles differ only in the $\pi/0$ entries of Table~\ref{tab:grover_phi} for $q_2, q_3, q_6, q_7$.

\subsection{\texorpdfstring{Actual Measurement Angle $\phi'_i$ with $X$ and $Z$ Corrections}{Actual Measurement Angle phi-prime i with X and Z Corrections}}
\label{sec:phi_prime}

The logical angles $\phi_i$ defined above fix the measurement \emph{basis} but not the \emph{adaptive} measurement schedule: each $\phi_i$ must be shifted by the byproduct operators accumulated along the flow before qubit $i$ is measured, as derived in Section~\ref{sec:flow}. We now specialize the general correction formula (Eq.~\eqref{phi2}) to the $2\times 9$ grid.

The grid admits a causal flow in the sense of Danos--Kashefi~\cite{Danos2007}, and one valid choice is the forward flow $f(i) = i + 2$, so the unique flow-predecessor of $i$ is $f^{-1}(i) = i-2$. Each measurement outcome $s_j$ propagates forward under this flow as an $X$-byproduct at $f(j)$ and a $Z$-byproduct at every other graph-neighbor of $f(j)$; collecting these contributions at the receiver $i$ gives the dependency sets of Section~\ref{sec:flow},
\begin{itemize}
    \item $S_X(i) = \{\, j : f(j) = i \,\} = \{i-2\}$: the $X$-byproduct at qubit $i$ is inherited from its unique flow-predecessor;
    \item $S_Z(i) = \{\, j : i \in N_G(f(j)) \setminus\{j\}\,\}$: the $Z$-byproduct at qubit $i$ is accumulated from every earlier measurement $j$ whose flow-image $f(j)$ is graph-adjacent to $i$.
\end{itemize}

Let $VCZ = \{4, 5, 14, 15\}$ denote the indices at which vertical $CZ$ edges are placed. On our grid, the condition $i \in N_G(f(j))\setminus\{j\}$ with $f(j)=j+2$ admits two types of solutions: (a) the \emph{horizontal} case $i = j+4$, yielding $j = i-4$, which contributes whenever $i-4$ exists; and (b) a \emph{cross-rail} case in which $f(j) = j+2 \in VCZ$ and $i$ is the vertical partner of $j+2$, which requires $i \in VCZ$. The two cases have complementary physical origins. The horizontal contribution is the usual chain propagation. The cross-rail contribution is a direct consequence of the commutation identity
\begin{equation}
    CZ\,(I \otimes X) = (Z \otimes X)\,CZ,
    \label{eq:cz_commutation}
\end{equation}
which states that an $X$-byproduct arriving at one endpoint of a vertical $CZ$ edge is converted into an additional $Z$-byproduct at the opposite endpoint when the entangling gate commutes past it. Consequently, whenever the receiving qubit $i$ itself lies on a vertical $CZ$ edge ($i \in VCZ$), its $Z$-correction set picks up the extra index corresponding to the flow-predecessor on the opposite rail. Enumerating both cases, the correction sets take the closed form
\begin{equation}
\begin{aligned}
S_X(i) &=
    \begin{cases}
        \emptyset & i \in \{0,1\}, \\
        \{i-2\} & i \in \{2,3,\dots,17\},
    \end{cases} \\[4pt]
S_Z(i) &=
    \begin{cases}
        \emptyset & i \in \{0,1,2,3\}, \\
        \{i-4\} & i \in \{4,\dots,17\} \setminus VCZ, \\
        \{i-4,\; i-1\} & i \in \{4, 14\}, \\
        \{i-4,\; i-3\} & i \in \{5, 15\}.
    \end{cases}
\end{aligned}
\label{eq:correction_sets}
\end{equation}
The two-element cases for $i\in\{4,5,14,15\}$ are exactly the CZ-commutation contributions of Eq.~\eqref{eq:cz_commutation}.

Substituting these $S_X(i), S_Z(i)$ into the general correction formula Eq.~\eqref{phi2} yields the adaptive angles $\phi'_i$ which, by construction of the flow, absorb all accumulated byproduct operators into the measurement basis in real time; the output distribution on $(q_{16}, q_{17})$ is therefore deterministic on every shot (numerical verification is reported in Section~\ref{sec:grover_impl}).

\subsection{Implementation and Results}
\label{sec:grover_impl}

Algorithm~\ref{alg:InitState} implements the resource-state preparation for the $2\times 9$ grid of Fig.~\ref{fig:MBQC_Grover}. Since Qiskit initializes every qubit in $\ket{0}$, a Hadamard layer is applied to reach the product state $\ket{+}^{\otimes 18}$, after which $CZ$ gates are applied along every horizontal edge in $E_{\text{horiz}}$ and along the two vertical edges in $E_{\text{vert}}$ indexed by $VCZ = \{4,5,14,15\}$ (Eq.~\eqref{eq:grover_edges}).

\begin{algorithm}
\caption{MBQC resource-state preparation}\label{alg:InitState}
\begin{algorithmic}
    \State \textbf{Input:} circuit, the number of qubits \( n \), \( VCZ \)
    \State \textbf{Output:} circuit

    \For{$i$ in $[0,n-1]$}
        \State circuit.h($i$)
    \EndFor

    \State $j=n/2$

    \For{$i$ in $[0,j-2]$}
        \State circuit.cz($2i, 2i+2$)
        \State circuit.cz($2i+1, 2i+3$)
    \EndFor

    \For{$i$ in $VCZ$}
         \If{$i\mod2 = 0$}
            \State circuit.cz($i, i+1$)
        \EndIf
    \EndFor
    \State \textbf{return} circuit
\end{algorithmic}
\end{algorithm}

Algorithm~\ref{alg:Measurement} realizes the adaptive measurement $M^{\phi'_i}$ of Eq.~\eqref{Mphi} on a circuit-based backend. Dynamic feed-forward is implemented through Qiskit's classically-conditioned gate \texttt{circuit.gate($i$).c\_if($j,1$)}, which applies \texttt{gate} to qubit $i$ if and only if the prior measurement outcome stored in classical register~$j$ equals~$1$. Following the dependency sets of Eq.~\eqref{eq:correction_sets}, the algorithm first applies all $Z$-corrections (horizontal contribution from $s_{i-4}$, plus a cross-rail contribution $s_{i-1}$ or $s_{i-3}$ when $i\in VCZ$), then the $X$-correction from $s_{i-2}$, and finally the $R_Z(-\phi_i)H$ pre-rotation followed by a $Z$-basis measurement, which together simulate $M^{\phi'_i}$ via the identity of Eq.~\eqref{MphiMz}. The XOR-parity sums $\bigoplus_{j\in S_X(i)} s_j$ and $\bigoplus_{j\in S_Z(i)} s_j$ of Eq.~\eqref{phi2} are realized by issuing one \texttt{c\_if}-conditioned gate per dependency rather than computing the parity on the classical register first: because $X^2 = Z^2 = I$, the composed effect on the target qubit is exactly $X^{\bigoplus s_j}$ (resp.\ $Z^{\bigoplus s_j}$), as introduced in Section~\ref{sec:meas_sim}.

\begin{algorithm}
\caption{Flow-corrected measurement \(M^{\phi'}\)}\label{alg:Measurement}
\begin{algorithmic}
    \State \textbf{Input:} circuit, qubit index \( i \), initial angle \( \phi \), \( VCZ \)
    \State \textbf{Output:} circuit

    \State \(\triangleright\)~\textit{Z-correction}
    \If{$i-4 \geq 0$}
        \State circuit.z($i$).c\_if($i-4,1$)
    \EndIf

    \If{$i \in VCZ$ and $i\mod 2 =0 $ and  $i-1 \geq 0$}
        \State circuit.z($i$).c\_if($i-1,1$)
    \EndIf

    \If{$i \in VCZ$ and $i\mod 2 = 1 $ and  $i-3 \geq 0$}
        \State circuit.z($i$).c\_if($i-3,1$)
    \EndIf

    \State \(\triangleright\)~\textit{X-correction}
    \If{$i-2 \geq 0$}
        \State circuit.x($i$).c\_if($i-2,1$)
    \EndIf

    \State \(\triangleright\)~\textit{$M_i^{\phi_i}$ measurement}
    \State circuit.rz($-\phi,i$)
    \State circuit.h($i$)
    \State circuit.measure($i,i$)
    \State \textbf{return} circuit
\end{algorithmic}
\end{algorithm}

Algorithm~\ref{alg:Main} assembles the complete MBQC two-qubit Grover circuit by composing Algorithms~\ref{alg:InitState} and~\ref{alg:Measurement}. After resource-state preparation with $n = 18$ qubits and $VCZ = \{4, 5, 14, 15\}$, the logical angles $\phi_i$ are assigned according to the oracle bits $(v_0, v_1)$ through the $w_x = 1 - v_x$ rule of Table~\ref{tab:grover_phi}, and the 16 non-output qubits $q_0, \dots, q_{15}$ are measured sequentially in ascending index order using Alg.~\ref{alg:Measurement}. On the two output qubits $q_{16}, q_{17}$ only the $X$-corrections are applied before the final $Z$-basis readout: since $Z$-corrections commute with $Z$-basis measurement (a $Z$-byproduct on a qubit about to be measured in the $Z$-basis leaves the outcome distribution invariant up to a global phase), they can be omitted without affecting correctness. The resulting transpiled Qiskit dynamic circuit for oracle~=~``00'' spans all 18 qubits and is, for legibility, displayed in landscape orientation as Fig.~\ref{fig:MBQC2QGrover} of Appendix~\ref{app:qiskit_grover}.

\begin{figure*}[t]
    \centering
    \includegraphics[width=0.9\textwidth]{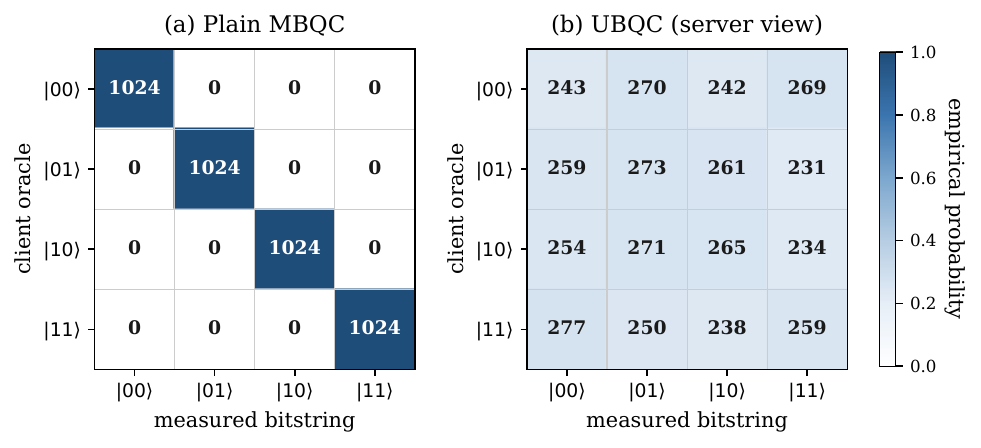}
    \caption{Outcome heatmaps for the 18-qubit Grover pattern of Fig.~\ref{fig:MBQC_Grover}, aggregated over $1024$ shots per oracle. Rows denote the client's oracle and columns use the Qiskit output bitstring order $q_{17}q_{16}$. \textbf{(a) Plain MBQC}: each row is diagonal with $1024/1024$ shots on the marked state. \textbf{(b) UBQC server view}: suppressing the final output unblinding yields an approximately uniform row, consistent with the angle/outcome-hiding analysis of Section~\ref{sec:ubqc_security}.}
    \label{fig:GroverOutcomes}
\end{figure*}

All simulations were executed on Qiskit Aer's \texttt{AerSimulator} with the default statevector method, and each of the four oracle instances was sampled over $1024$ shots. Unless otherwise stated, all experiments in this paper were performed with Qiskit~$\geq 1.0$ and \texttt{qiskit-aer}~$\geq 0.14$ on Python~$3.10$; the UBQC pad arrays $\{\theta_i\}_{i=0}^{17}$ and $\{r_i\}_{i=0}^{17}$ used to generate Fig.~\ref{fig:GroverOutcomes}(b) and Fig.~\ref{fig:noise_comparison} were drawn with \texttt{numpy.random.default\_rng(seed=42)}, and the supplementary package provides the source notebooks, a canonical histogram-count table, the final review figures, and a minimal execution environment description to support independent reproduction. The measurement outcomes on the output qubits $(q_{16}, q_{17})$, reported in the Qiskit bitstring order $q_{17}q_{16}$, are shown in Fig.~\ref{fig:GroverOutcomes}(a): every shot lands on the marked bitstring for all four oracles, giving a total $4096/4096$ success rate. This is the expected ceiling rather than a statistical coincidence: the circuit-model two-qubit Grover deterministically outputs the marked bitstring in a single iteration (Section~\ref{sec:grover_circuit}), and the flow-induced correction schedule of Eq.~\eqref{eq:correction_sets} absorbs every measurement-dependent byproduct into $\phi'_i$, so under noise-free execution the MBQC output distribution inherits this determinism exactly. The $4096/4096$ rate thus validates that the \texttt{c\_if}-based feed-forward of Alg.~\ref{alg:Measurement} correctly realizes this absorption and serves as the noise-free deterministic baseline that Section~\ref{sec:robustness} later perturbs with depolarizing noise.


\begin{algorithm}
\caption{Plain MBQC Grover circuit}\label{alg:Main}
\begin{algorithmic}
    \State \textbf{Input:} the number of qubits $n=18$, $VCZ = \{ 4,5,14,15 \}$, oracle
    \State \textbf{Output:} circuit

    \State \(\triangleright\)~\textit{initial setting}
    \State circuit = Initialize\_quantum\_circuit(n,n)

    \State MBQC\_Initial\_State\_Set(circuit,$n,VCZ$) \hfill (Alg.~\ref{alg:InitState})

    \State \(\triangleright\)~\textit{set $\{\phi_i\}$}
    \State $\phi_i = 0$ for $i\in[0,n-1]\setminus [10,11]$
    \State $\phi_i = \pi$ for $i\in[10,11]$
    \If{oracle = ``00'' or ``10''}
        \State $\phi_i = \pi$ for $i \in \{2,6\}$
    \EndIf
    \If{oracle = ``00'' or ``01''}
        \State $\phi_i = \pi$ for $i \in \{3,7\}$
    \EndIf

    \State \(\triangleright\)~\textit{intermediate-qubit measurements}
    \For{$i$ in $[0,n-3]$}
        \State $M^{\phi'}(circuit, i, \phi_i, VCZ)$ \hfill (Alg.~\ref{alg:Measurement})
    \EndFor

    \State \(\triangleright\)~\textit{output-qubit readout with $X$-correction}
    \For{$i$ in $[n-2,n-1]$}

    \If{$i-2 \geq 0$}
        \State circuit.x($i$).c\_if($i-2,1$)
    \EndIf

    \State circuit.measure($i,i$)

    \EndFor
    \State \textbf{return} circuit

\end{algorithmic}
\end{algorithm}

\section{UBQC Protocol on MBQC Two-Qubit Grover Simulation}
\label{sec:ubqc}

This section combines the MBQC pattern of Section~\ref{sec:grover} with the UBQC protocol of Broadbent, Fitzsimons, and Kashefi~\cite{Broadbent2009} to study fixed-graph angle-based blindness for delegated two-qubit Grover search. Section~\ref{sec:ubqc_protocol} briefly reviews the protocol in its original brickwork formulation. Section~\ref{sec:ubqc_grover} then states the two simulation-level departures used here and the resulting circuit-level recipe, Section~\ref{sec:ubqc_results} presents the Qiskit implementation and outputs, and Section~\ref{sec:ubqc_security} analyses what remains blind and what does not.

We stress at the outset that our instantiation differs from the original UBQC of~\cite{Broadbent2009} in its graph resource: the brickwork state $\mc{G}_{n,m}$ is replaced by the algorithm-specific $2\times 9$ flow of Section~\ref{sec:grover}, a substitution motivated by the qubit-count ceiling of current simulators. Consequently, the blindness claim established in Section~\ref{sec:ubqc_security} is restricted to the measurement angles $\{\phi_i\}$ and outcomes $\{s_i\}$; the graph topology itself remains visible to the server. Structural blindness---obtained for free by an algorithm-independent brickwork resource---is recovered only upon adopting that layout and is left as future work (Sections~\ref{sec:resource_overhead}, \ref{sec:conclusion}).

\subsection{UBQC Protocol}
\label{sec:ubqc_protocol}

This subsection reviews the Universal Blind Quantum Computation (UBQC) protocol of Broadbent, Fitzsimons, and Kashefi~\cite{Broadbent2009} in its original brickwork formulation, which serves as the conceptual basis for the $2\times 9$-grid instantiation developed in Section~\ref{sec:ubqc_grover}.

The protocol's security goal is \textit{blindness}: for a fixed graph size, Bob's classical view must be identically distributed across all admissible computations, so that it carries no information about Alice's input, output, or measurement pattern beyond the graph size itself (cf.~\cite[Def.~2]{Broadbent2009}); graph-size leakage is explicitly permitted by the formal definition.

On a brickwork graph $\mc{G}_{n,m}$---an $n\times m$ rectangular array of vertices with $n$ logical-qubit rows and $m$ columns proportional to the circuit's two-qubit-gate depth, in the sense of~\cite[Def.~4]{Broadbent2009}---each vertex $(x,y)$ is assigned a logical measurement angle $\phi_{x,y}$ together with flow-derived dependency sets $D_{x,y}$ ($X$-dependencies) and $D'_{x,y}$ ($Z$-dependencies), each a set of preceding vertices. Writing $s^X_{x,y} = \bigoplus_{(x',y')\in D_{x,y}} s_{x',y'}$ and $s^Z_{x,y} = \bigoplus_{(x',y')\in D'_{x,y}} s_{x',y'}$ for the corresponding measurement-outcome parities (with the convention that these parities vanish on the input column), the adaptive angle that absorbs the accumulated byproduct operators is
\begin{equation}
    \phi'_{x,y} = (-1)^{s^X_{x,y}} \phi_{x,y} + s^Z_{x,y}\, \pi,
    \label{eq:phi_prime_xy}
\end{equation}
which is the brickwork-indexed restatement of the general correction formula Eq.~\eqref{phi2} of Section~\ref{sec:flow}.

Blindness is superimposed on this MBQC layer through two independent per-qubit one-time pads. For every vertex $(x,y)$ Alice draws a \textit{basis pad} $\theta_{x,y} \in \{0, \pi/4, \dots, 7\pi/4\}$ and an \textit{outcome pad} $r_{x,y} \in \{0,1\}$ uniformly at random, and then sends Bob only the blinded angle
\begin{equation}
    \delta_{x,y} = \phi'_{x,y} + \theta_{x,y} + \pi\, r_{x,y} \pmod{2\pi}.
    \label{eq:blind_angle}
\end{equation}
The two pads play complementary roles. Because $\theta_{x,y}$ is uniform over the 8-element set, the transmitted angle $\delta_{x,y}$ is likewise uniformly distributed and information-theoretically independent of $\phi'_{x,y}$; this conceals the measurement axis and, hence, the computation. The outcome pad $r_{x,y}$ conceals the reported bit: the identity $\lvert \pm_{\varphi + \pi}\rangle = \lvert \mp_{\varphi}\rangle$ (up to a global phase) implies that shifting the measurement angle by $\pi r_{x,y}$ swaps Bob's $\lvert + \rangle / \lvert - \rangle$ outcome labels whenever $r_{x,y} = 1$, so Bob's reported bit $s_{x,y}$ differs from the logical outcome by $r_{x,y}$. Alice recovers the logical bit via the post-measurement flip $s'_{x,y} \leftarrow s_{x,y} \oplus r_{x,y}$ (final step of Protocol~\ref{prot:UBQC} below). The 8-element angle alphabet $\{k\pi/4\}_{k=0}^{7}$ is the natural resolution induced by Clifford$+T$ measurement patterns~\cite[Sec.~II]{Broadbent2009}; the two-qubit Grover instance considered in Section~\ref{sec:ubqc_grover} uses only the $\{0,\pi\}$ subset for $\phi_{x,y}$ (cf.~Section~\ref{sec:grover_phi}), but the full $\pi/4$-resolution pad is retained here to keep the protocol statement universal.

Protocol~\ref{prot:UBQC} below gives the full procedure; the Qiskit instantiation and the simulation-level simplifications it entails are deferred to Section~\ref{sec:ubqc_grover}.

\begin{protocol}
\caption{Universal Blind Quantum Computation (UBQC)~\cite{Broadbent2009}}
\label{prot:UBQC}
\begin{algorithmic}[1]
    \LongState{\textbf{Input:} brickwork dimensions $(n,m)$; Alice's logical measurement angles $\{\phi_{x,y}\}$ with flow-derived dependency sets $\{D_{x,y}, D'_{x,y}\}$.}
    \State \textbf{Output:} Alice's logical measurement outcomes $\{s'_{x,y}\}$.

    \Statex \textbf{\textit{Phase 1: Alice's state preparation}}
    \For{$x = 1, \dots, n$}
        \For{$y = 1, \dots, m$}
            \State Alice draws $\theta_{x,y} \in_R \{0, \pi/4, \dots, 7\pi/4\}$.
            \LongState{Alice prepares $\ket{\psi_{x,y}} = \ket{+_{\theta_{x,y}}}$ and sends it to Bob.}
        \EndFor
    \EndFor

    \Statex \textbf{\textit{Phase 2: Bob's entanglement}}
    \LongState{Bob applies $CZ$ between adjacent qubits to form the brickwork graph state $\mc{G}_{n,m}$.}

    \Statex \textbf{\textit{Phase 3: Interaction and measurement (in flow order)}}
    \For{$x = 1, \dots, n$}
        \For{$y = 1, \dots, m$}
            \LongState{Alice computes $\phi'_{x,y}$ from Eq.~\eqref{eq:phi_prime_xy}, using the convention that $s^X_{x',y'}$ and $s^Z_{x',y'}$ vanish on the input column.}
            \LongState{Alice draws $r_{x,y} \in_R \{0,1\}$ and sends $\delta_{x,y} = \phi'_{x,y} + \theta_{x,y} + \pi r_{x,y}$ to Bob.}
            \LongState{Bob measures qubit $(x,y)$ in the basis $\{\ket{+_{\delta_{x,y}}}, \ket{-_{\delta_{x,y}}}\}$ and returns $s_{x,y} \in \{0,1\}$ to Alice.}
            \State Alice sets $s'_{x,y} \leftarrow s_{x,y} \oplus r_{x,y}$.
        \EndFor
    \EndFor
\end{algorithmic}
\end{protocol}

\subsection{Two-Qubit Grover with UBQC Protocol}
\label{sec:ubqc_grover}

We now instantiate Protocol~\ref{prot:UBQC} on the $2\times 9$ MBQC pattern of Section~\ref{sec:grover}. Two instantiation-level adjustments to the original Broadbent--Fitzsimons--Kashefi formulation are worth stating up front; neither affects the angle-blindness guarantee of Section~\ref{sec:ubqc_protocol}, and both are chosen to isolate the in-circuit realization of that guarantee within current simulator capacity.

First, for simulation convenience we implement Alice and Bob within a single Qiskit circuit; the quantum-state and classical-bit exchange of the original protocol is absorbed into one program, with Alice's pads $(\theta_{x,y}, r_{x,y})$ represented as classical variables. This is a simulation-level encoding of the same angle, outcome, and feed-forward update rules; it preserves the distributional pad argument for the server-visible angles and outcomes, but should not be read as an implementation of a physically separated client and server.

\begin{sidewaysfigure*}[p]
    \centering
    \includegraphics[width=\textheight]{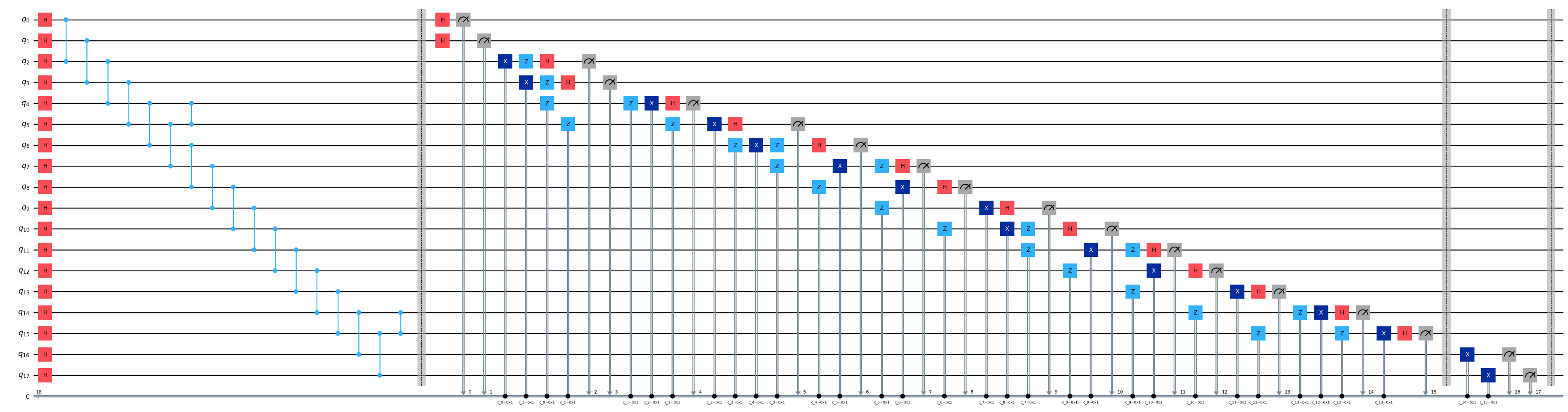}
\caption{Qiskit dynamic-circuit realization of the MBQC two-qubit Grover pattern of Fig.~\ref{fig:MBQC_Grover} (oracle = ``00''). Each measured qubit $q_i$ receives the \texttt{c\_if}-conditioned $Z$ and/or $X$ corrections derived from Eq.~\eqref{eq:correction_sets}, followed by an $R_Z(-\phi_i)H$ pre-rotation (Eq.~\eqref{MphiMz}) and a $Z$-basis readout; the output qubits $q_{16}, q_{17}$ receive only the final $X$-correction and are measured directly in the $Z$-basis.}
    \label{fig:MBQC2QGrover}
\end{sidewaysfigure*}

Second, we run the protocol on the $2\times 9$ MBQC grid directly rather than embedding it in a brickwork state, so Bob learns the graph shape in addition to its size. Pads $(\theta_{x,y}, r_{x,y})$ are still drawn for every vertex, so the measurement angles $\phi'_{x,y}$ themselves remain information-theoretically hidden; in particular, Alice's four-way oracle choice---encoded in the measurement angles of qubits $q_2, q_3, q_6, q_7$ in Fig.~\ref{fig:MBQC2QGrover}---is hidden from Bob through the angle/outcome transcript studied here. This inheritance is not accidental: the pad-induced uniformity of $\delta_{x,y}$ in Eq.~\eqref{eq:blind_angle} is a per-vertex property independent of the graph topology, so the angle-blindness argument of Section~\ref{sec:ubqc_protocol} carries over to the $2\times 9$ grid for fixed public topology, even though the full BFK09 blindness guarantee (which additionally conceals the graph topology) does not.

Concretely, three modifications to the MBQC simulation of Section~\ref{sec:grover} are required:
\begin{itemize}
    \item Alice prepares each qubit in $\ket{+_{\theta_{x,y}}}$ rather than $\ket{+}$, using the basis pad $\theta_{x,y}$.
    \item Each measurement uses the blinded angle $\delta_{x,y} = \phi'_{x,y} + \theta_{x,y} + \pi r_{x,y}$ instead of $\phi'_{x,y}$.
    \item The reported outcome $s_{x,y}$ is unblinded by $s'_{x,y} \leftarrow s_{x,y} \oplus r_{x,y}$, realized in-circuit (Section~\ref{sec:ubqc_meas}) by an $X$ gate and a re-measurement of qubit $(x,y)$ whenever $r_{x,y}=1$.
\end{itemize}
Bob's $CZ$ entanglement (Phase~2 of Protocol~\ref{prot:UBQC}) and the $\phi'$ feed-forward derived from the flow (Section~\ref{sec:flow}) are unchanged. Details of the first two modifications follow.

\subsubsection{Alice's Preparation in Qiskit}
\label{sec:ubqc_alice_prep}

The states $\{\ket{+_\theta}\}_{\theta\in\{k\pi/4\}_{k=0}^{7}}$ required by Phase~1 of Protocol~\ref{prot:UBQC} satisfy $\ket{+_{\theta}} = R_Z(\theta)\ket{+}$ up to a global phase, so Alice's preparation on qubit $(x,y)$ reduces to a Hadamard followed by $R_Z(\theta_{x,y})$ on a qubit initialized to $\ket{0}$. Consistent with the single-circuit convention stated above, the $\theta_{x,y}$ draws are generated inside the Qiskit program and the corresponding $R_Z$ rotations are inserted directly into the circuit, skipping the explicit Alice-to-Bob state transmission.

\subsubsection{\texorpdfstring{Measurement Method for UBQC, $M^{\delta}$}{Measurement Method for UBQC, M(delta)}}
\label{sec:ubqc_meas}

Bob measures qubit $(x,y)$ in the basis $\{\ket{+_{\delta_{x,y}}}, \ket{-_{\delta_{x,y}}}\}$ with $\delta_{x,y} = \phi'_{x,y} + \theta_{x,y} + \pi r_{x,y}$. Because $R_Z(-\theta)\ket{\pm_{\phi+\theta}} = \ket{\pm_\phi}$ up to a global phase, any $\{\ket{+_{\phi+\theta}}, \ket{-_{\phi+\theta}}\}$ measurement can be realized as a pre-rotation $R_Z(-\theta)$ followed by a $\{\ket{+_\phi}, \ket{-_\phi}\}$ measurement, i.e.\ $M_{x,y}^{\phi + \theta} = M_{x,y}^{\phi}\, R_Z(-\theta)$. Substituting $\theta \to \theta_{x,y} + \pi r_{x,y}$ and using the additivity $R_Z(-\alpha)R_Z(-\beta) = R_Z(-(\alpha+\beta))$ gives
\begin{equation}\label{eq:blind_delta_measure}
M_{x,y}^{\delta_{x,y}} = M_{x,y}^{\phi'_{x,y}}\, R_Z(-\theta_{x,y})\, R_Z(-\pi r_{x,y}),
\end{equation}
so in Qiskit the blinded measurement is implemented by prepending $R_Z(-\theta_{x,y})\, R_Z(-\pi r_{x,y})$ to the $M_{x,y}^{\phi'_{x,y}}$ primitive of Section~\ref{sec:grover}. The outcome $s_{x,y}$ is then unblinded by $s'_{x,y} \leftarrow s_{x,y} \oplus r_{x,y}$, which Section~\ref{sec:ubqc_results} realizes at the circuit level by applying an $X$ gate and re-measuring qubit $(x,y)$ whenever $r_{x,y}=1$; this reproduces Alice's final step in Protocol~\ref{prot:UBQC}.

\subsection{Implementation and Results}
\label{sec:ubqc_results}

For the Qiskit implementation we adopt the linear qubit indexing $i \in \{0, \dots, 17\}$ of Section~\ref{sec:grover} in place of the brickwork double index $(x,y)$ used above; the per-vertex pads $\theta_i, r_i$ and angles $\phi'_i, \delta_i$ correspond term-by-term to their $(x,y)$-indexed counterparts.

Algorithm~\ref{alg:BlindInitState} realizes Phase~1 of Protocol~\ref{prot:UBQC}: each qubit receives the pad-dependent initial state $\ket{+_{\theta_i}}$, prepared by a Hadamard followed by $R_Z(\theta_i)$ with $\theta_i \in_R \{0, \pi/4, \dots, 7\pi/4\}$. The subsequent $CZ$ entanglement pattern---horizontal edges on consecutive qubits plus vertical edges on $VCZ=\{4,5,14,15\}$---is identical to Algorithm~\ref{alg:InitState}.

\begin{algorithm}
\caption{Blind MBQC resource-state preparation}\label{alg:BlindInitState}
\begin{algorithmic}
    \State \textbf{Input:} circuit, $n$, $\{\theta_{i}\}_{i\in[0,n-1]}$, $VCZ$
    \State \textbf{Output:} circuit

    \For{$i$ in $[0,n-1]$}
        \State circuit.h($i$)
        \State circuit.rz($\theta_i,i$)
    \EndFor

    \State $j=n/2$

    \For{$i$ in $[0,j-2]$}
        \State circuit.cz($2i, 2i+2$)
        \State circuit.cz($2i+1, 2i+3$)
    \EndFor

    \For{$i$ in $VCZ$}
         \If{$i\mod2 = 0$}
            \State circuit.cz($i, i+1$)
        \EndIf
    \EndFor
    \State \textbf{return} circuit
\end{algorithmic}
\end{algorithm}

Algorithm~\ref{alg:BlindMeasurement} realizes the blinded measurement $M_i^{\delta_i}$ of Eq.~\eqref{eq:blind_delta_measure}. The $R_Z(-\pi r_i)$ factor is implemented as a $Z$ gate when $r_i = 1$ (coinciding with $R_Z(-\pi)$ up to a global phase); a subsequent $R_Z(-\theta_i)$ rotation and the unblinded measurement primitive $M_i^{\phi'_i}$ of Algorithm~\ref{alg:Measurement} complete the realization of $M_i^{\delta_i}$.

\begin{algorithm}
\caption{Blinded measurement $M^{\delta}$}\label{alg:BlindMeasurement}
\begin{algorithmic}
    \State \textbf{Input:} circuit, qubit index $i$, $\theta_{i}$, $r_{i}$, $\phi_i$, $VCZ$
    \State \textbf{Output:} circuit

    \State \(\triangleright\)~\textit{$\pi r_i$-Rotation}
    \If{$r_i = 1$}
        \State circuit.z($i$)
    \EndIf

    \State \(\triangleright\)~\textit{$(-\theta_i)$-Rotation}
    \State circuit.rz($-\theta_i,i$)

    \State \(\triangleright\)~\textit{$M_i^{\phi'}$ measurement}
    \State $M^{\phi'}(circuit, i, \phi_i, VCZ)$ \hfill (Alg.~\ref{alg:Measurement})

    \State \textbf{return} circuit

\end{algorithmic}
\end{algorithm}

Algorithm~\ref{alg:BlindMain} assembles the full simulation on $n=18$ qubits with $VCZ = \{4,5,14,15\}$ and the oracle-dependent angle assignment $\{\phi_i\}$ of Algorithm~\ref{alg:Main}. After drawing the pads $\{\theta_i\}$ and $\{r_i\}$, it initializes the state via Algorithm~\ref{alg:BlindInitState}, applies $M_i^{\delta_i}$ of Algorithm~\ref{alg:BlindMeasurement} to every input and intermediate qubit, and finally measures the output qubits $q_{16}$ and $q_{17}$ with the same $X$-correction-via-\texttt{c\_if} procedure as Algorithm~\ref{alg:Main}.

The post-measurement flip $s'_i \leftarrow s_i \oplus r_i$ of Section~\ref{sec:ubqc_meas} is executed in-circuit: whenever $r_i = 1$, an $X$ gate is applied to qubit $i$ immediately after its measurement and the qubit is re-measured, overwriting the previously stored bit. This keeps the downstream \texttt{c\_if} corrections of Section~\ref{sec:grover} synchronized with the unblinded outcomes without requiring runtime classical arithmetic on the register. The output qubits $q_{16}, q_{17}$ carry only a global phase from their own $R_Z(\theta)$ pads---since the flow of Fig.~\ref{fig:MBQC_Grover} projects them onto $Z$-eigenstates before readout---so no explicit $\theta$-unblinding is required on the output register.

The circuit produced by Algorithm~\ref{alg:BlindMain} (with oracle ``00'') is displayed in landscape orientation as Fig.~\ref{fig:BlindMBQC2QGrover} of Appendix~\ref{app:blind_qiskit_grover}. Its output distribution (the client's view) reproduces the correct Grover statistics of Fig.~\ref{fig:GroverOutcomes}(a) for all four oracles, achieving the same $4096/4096$ deterministic rate as the non-blind run of Section~\ref{sec:grover_impl}; the complementary server view, obtained by suppressing the $s_{14}, s_{15}$ unblinding flips that drive the final \texttt{c\_if} $X$-correction on $q_{16}, q_{17}$, is shown in Fig.~\ref{fig:GroverOutcomes}(b) and used in Section~\ref{sec:ubqc_security} as a simulation-level consistency check of the output-hiding behavior.

\begin{algorithm}
\caption{Blind MBQC Grover circuit}\label{alg:BlindMain}
\begin{algorithmic}
    \State \textbf{Input:} the number of qubits $n=18$, $VCZ = \{ 4,5,14,15 \}$, oracle
    \State \textbf{Output:} circuit

    \State \(\triangleright\)~\textit{generate random array}
    \State $\theta_i \in_R \{0,\pi/4...,7\pi/4\}$ for $i\in[0,n-1]$
    \State $r_i \in_R \{0,1\}$ for $i\in[0,n-1]$

    \State \(\triangleright\)~\textit{initial setting}
    \State circuit = Initialize\_quantum\_circuit(n,n)
    \State Blind\_Initial\_State\_Set(circuit,$n,\{\theta_i\},VCZ$) \hfill (Alg.~\ref{alg:BlindInitState})

    \State \(\triangleright\)~\textit{set $\{\phi_i\}$}
    \State $\phi_i = 0$ for $i\in[0,n-1]\setminus [10,11]$
    \State $\phi_i = \pi$ for $i\in[10,11]$
    \If{oracle = ``00'' or ``10''}
        \State $\phi_i = \pi$ for $i \in \{2,6\}$
    \EndIf
    \If{oracle = ``00'' or ``01''}
        \State $\phi_i = \pi$ for $i \in \{3,7\}$
    \EndIf

    \State \(\triangleright\)~\textit{intermediate-qubit measurements}
    \For{$i$ in $[0,n-3]$}
        \State $M^{\delta}(circuit, i,\theta_i, r_i, \phi_i, VCZ)$ \hfill (Alg.~\ref{alg:BlindMeasurement})
        \If{$r_i = 1$}
            \State circuit.x($i$)
            \State circuit.measure($i,i$)
        \EndIf
    \EndFor

    \State \(\triangleright\)~\textit{output-qubit readout with $X$-correction}
    \For{$i$ in $[n-2,n-1]$}

    \If{$i-2 \geq 0$}
        \State circuit.x($i$).c\_if($i-2,1$)
    \EndIf

    \State circuit.measure($i,i$)

    \EndFor

    \State \textbf{return} circuit

\end{algorithmic}
\end{algorithm}



\subsection{Brickwork-State Universal-Gate Fragment}
\label{sec:brickwork_fragment}

The Grover experiment above deliberately uses a compact, algorithm-specific $2\times 9$ flow to keep the full dynamic-circuit simulation within the statevector budget. To check that the same Qiskit correction and blinding machinery is not limited to that hand-tailored flow, we additionally simulate a standalone $G_{2,5}$ brickwork-state fragment. The fragment has two rows and five columns, inputs on $q_0,q_1$, outputs on $q_8,q_9$, horizontal row-wise flow $f(i)=i+2$, and the brickwork vertical edges at $(q_4,q_5)$ and $(q_8,q_9)$. The measured vertices are $q_0,\dots,q_7$; all measurement angles below are written as integer indices $k$ denoting $\phi=k\pi/4$.

The four measurement patterns used in this check are collected in Appendix~\ref{app:brickwork_patterns} (Fig.~\ref{fig:brickwork_gate_patterns}). The angle lists are compiled from the brickwork-state measurement calculus of BFK09 under the qubit ordering, output convention, and adaptive-measurement rule of this Qiskit implementation; they are not copied as literal table entries from~\cite{Broadbent2009}. The selected cases cover identity padding, a single-qubit Clifford, an entangling primitive, and a non-Clifford/Clifford combination:
\begin{equation*}
\begin{array}{ll}
\mathrm{A}: I\otimes I, &
\mathrm{B}: H\otimes I,\\
\mathrm{C}: (I\otimes H)CZ, &
\mathrm{D}: T\otimes H.
\end{array}
\end{equation*}

\begin{table*}[tbp]
\centering
\caption{Brickwork-fragment gate cases. Angle entries are the indices $k$ in $\phi=k\pi/4$ for $q_0,\dots,q_7$. The readout column lists the verification input and output basis used in Fig.~\ref{fig:brickwork_client_server}.}
\label{tab:brickwork_cases}
\renewcommand{\arraystretch}{1.18}
\footnotesize
\setlength{\tabcolsep}{3pt}
\begin{tabular}{@{}l l p{4.0cm} p{3.8cm} l@{}}
\hline
Case & Implemented logical gate & $\phi$ pattern on $q_0,\dots,q_7$ & Verification input/readout & Client peak \\
\hline
A & $I\otimes I$ & $[0,0,0,0,0,0,0,0]$ & input $\ket{00}$, $Z/Z$ readout & $\ket{00}$ \\
B & $H\otimes I$ & $[2,0,2,0,2,0,0,0]$ & input $\ket{-0}$, $Z/Z$ readout & $\ket{10}$ \\
C & $(I\otimes H)CZ$ & $[0,0,0,2,6,2,0,2]$ & input $\ket{1+}$, $Z/Z$ readout & $\ket{11}$ \\
D & $T\otimes H$ & $[0,2,4,2,1,6,4,0]$ & input $\ket{+-}$, upper $X(\pi/4)$ and lower $Z$ readout & $\ket{01}$ \\
\hline
\end{tabular}
\end{table*}

Case~C is intentionally reported as $(I\otimes H)CZ$ rather than as a bare $CZ$. In this five-column brickwork fragment, the lower output wire carries a deterministic local Clifford padding induced by the teleportation steps of the measurement pattern. This padding is not a correctness error: applying the known lower-wire $H$ frame to the output gives an exact $CZ$ operation, and in a larger brickwork computation the same local Clifford can be absorbed into the next single-qubit layer or into the final readout frame. We therefore describe Case~C as a $CZ$-equivalent entangling primitive up to a known local Clifford frame, not as a direct bare-$CZ$ realization by the isolated fragment.

\begin{figure*}[tbp]
    \centering
    \includegraphics[width=0.95\textwidth]{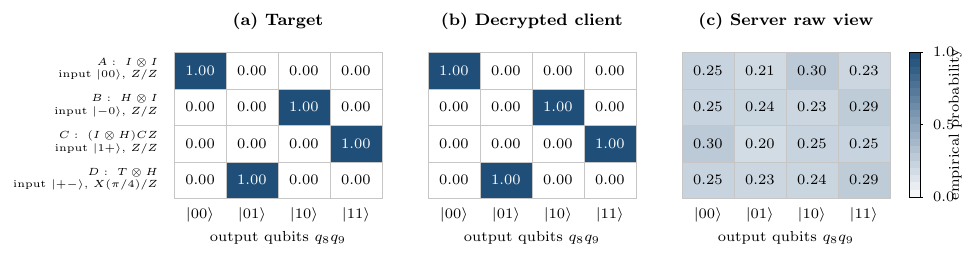}
    \caption{Target, client, and server views for the $G_{2,5}$ brickwork universal-gate fragment. The client-decrypted view matches the target logical output selected in Table~\ref{tab:brickwork_cases}, while the server raw view is uniform-like because the final output one-time pad is hidden.}
    \label{fig:brickwork_client_server}
\end{figure*}

We then apply the same UBQC-style pads used in Section~\ref{sec:ubqc_results}. For each case, the client view removes the output $\theta$ pad, applies the flow-induced output correction, and decrypts the final output one-time pad. The server view keeps the raw output bits before this client-side decryption. Figure~\ref{fig:brickwork_client_server} compares the target logical distribution, the client-decrypted distribution, and the server raw distribution. The client panel reproduces the target peak for all four cases in Table~\ref{tab:brickwork_cases}; the server panel is close to uniform over the four output bitstrings after averaging over balanced output pads. This experiment is a fragment-level check of universal brickwork primitives under the same dynamic-circuit UBQC machinery, rather than a claim of full algorithmic structural blindness.

\subsection{Security Analysis and Limitations}
\label{sec:ubqc_security}

Our main Grover experiment uses a custom $2 \times 9$ grid graph state in place of a full algorithmic brickwork state---a deliberate scope choice that isolates the angle-blindness mechanism within the qubit budget of current statevector simulators (Section~\ref{sec:comparison}). The $G_{2,5}$ brickwork experiment of Section~\ref{sec:brickwork_fragment} checks representative universal-gate primitives, but it is still a fragment-level experiment rather than a complete structurally blind Grover embedding. We therefore give a careful account of which aspects of the computation remain blind and which do not. We distinguish two regimes: the security of the \textit{oracle information}, and the privacy of the \textit{algorithm structure}.

\vspace{0.2cm}

\noindent \textbf{Information Security of the Oracle.}
The primary goal of the blinded Grover's algorithm is to conceal the marked state (the oracle) from the server; in our implementation that choice is encoded in the measurement angles of $q_2, q_3, q_6, q_7$ via the $w_x = 1 - v_x$ rule of Table~\ref{tab:grover_phi}. We state the resulting blindness claim precisely as a statement about the server's transcript. Let $\mathcal{T}_{\mathrm{srv}} = (\{\delta_i\}, \{s_i\}, G, E_{CZ})$ denote everything the server observes across a full run---the blinded angles $\delta_i$ it receives and raw measurement bits $s_i$ it returns for the measured non-output vertices, the graph $G$, and the $CZ$-edge pattern $E_{CZ}$---and let $\Phi = \{\phi'_i\}$ denote the unblinded angle pattern that encodes the oracle. Two elementary facts, both derived in Appendix~\ref{app:blind_itsec}, suffice to control what $\Phi$ leaks into $\mathcal{T}_{\mathrm{srv}}$.

\emph{Lemma~1 (angle pad).} For every non-output vertex $i$, the blinded angle $\delta_i = \phi'_i + \theta_i + \pi r_i \pmod{2\pi}$ is uniform on $\{k\pi/4\}_{k=0}^{7}$ and statistically independent of $\phi'_i$.

\emph{Lemma~2 (outcome pad).} Conditioned on any history of prior measurement outcomes, the server-visible bit $s_i$ is uniform on $\{0,1\}$ and independent of the unblinded outcome $s'_i = s_i \oplus r_i$, because $r_i \in_R \{0,1\}$ is fresh and private.

Taken together, the pair $(\delta_i, s_i)$ carries zero mutual information with $\phi'_i$ for every $i$, so $I\!\left(\Phi\, ; \mathcal{T}_{\mathrm{srv}} \setminus (G, E_{CZ})\right) = 0$. Conditional on the graph topology, which is public by construction, the server therefore cannot distinguish which of the four oracles $\{\text{``00'',``01'',``10'',``11''}\}$ was delegated from the blinded angles and reported outcomes alone. This is the standard BFK09 one-time-pad mechanism specialized to our fixed $2\times 9$ instance; the complementary leakage channel through $(G, E_{CZ})$ is analysed in the ``Leakage of Algorithm Structure'' paragraph below.

\vspace{0.2cm}

\noindent \textbf{Empirical Consistency Check.}
Lemmas~1--2 are an information-theoretic argument about the protocol, not a statement that our particular Qiskit implementation faithfully realises it. To sanity-check the latter, we compare two views of the \emph{same} simulation run, differing only in whether the output-stage unblinding flips are applied at post-readout. In the \textit{client view}, the in-circuit flips $s'_i \leftarrow s_i \oplus r_i$ of Algorithm~\ref{alg:BlindMain} are applied as specified, and the output distributions reproduce those of Fig.~\ref{fig:GroverOutcomes}(a) for all four oracles. In the \textit{server view}, the flips on $s_{14}$ and $s_{15}$---which alone determine the final \texttt{c\_if} $X$-correction on the output qubits $q_{16}, q_{17}$---are suppressed. Because the upstream MBQC corrections have already consumed unblinded outcomes in-circuit, the output stage is the only remaining point at which the pads $(r_{14}, r_{15})$ enter the server's view; denying those two pads therefore isolates the output-register realisation of Lemma~2. The resulting distribution is uniform over $\{\ket{00}, \ket{01}, \ket{10}, \ket{11}\}$ for every oracle (Fig.~\ref{fig:GroverOutcomes}(b)), the behaviour predicted by Lemma~2 and inconsistent with any alternative in which leaked pads bias the outcomes. We emphasise that the formal angle/outcome hiding statement rests on Lemmas~1--2 of Appendix~\ref{app:blind_itsec}; Fig.~\ref{fig:GroverOutcomes}(b) corroborates that our Qiskit implementation is consistent with those lemmas at the readout stage.

\vspace{0.2cm}

\noindent \textbf{Leakage of Algorithm Structure.}
By contrast, the \textit{structure} of the algorithm is not fully concealed. Standard UBQC uses a uniform brickwork state to hide gate placement, so that all computations appear topologically identical to the server. Our customized grid instead pins the vertical $CZ$ edges to the set $VCZ = \{4,5,14,15\}$, which encodes the two-qubit gate pattern specific to Grover's algorithm. A server with knowledge of the graph layout can therefore infer the location and sequence of the entangling gates and identify the computation as a Grover-style search.

\vspace{0.2cm}

\noindent \textbf{Resource Constraints.}
This structural leakage is a trade-off imposed by the simulation environment. A structurally blind brickwork instantiation of the two-qubit Grover computation would require an entangled resource substantially larger than the 18 qubits used here: the BFK09 brickwork width $m = 4(N_{2q}+k)+1$ (detailed in Section~\ref{sec:resource_overhead}) already places the two-qubit Grover instance at $n \times m = 26$ qubits for $k=1$ and $34$ for $k=2$, exceeding the practical capacity of current 20--30 qubit simulators for this specific demonstration. The $G_{2,5}$ fragment of Section~\ref{sec:brickwork_fragment} partially addresses this limitation by verifying the same dynamic-circuit machinery on brickwork universal-gate primitives, but it does not conceal the structure of the full Grover algorithm. Our work therefore focuses on validating the mechanism of angle-based blindness within these constraints.

\section{Discussion}
\label{sec:discussion}

\subsection{Robustness under Depolarizing Noise}
\label{sec:robustness}

To evaluate the noise robustness of our approach, we performed simulations under a depolarizing noise model and compared three scenarios: (1) the standard gate-based implementation (2 qubits, shallow-depth Grover circuit), (2) our plain MBQC simulation (18 qubits, Algorithm~\ref{alg:Main}), and (3) our UBQC-augmented MBQC simulation (also 18 qubits, with the per-qubit pads and in-circuit unblinding of Section~\ref{sec:ubqc_results}).

Depolarizing noise is inserted per gate, with single-qubit rate $p$ on every single-qubit gate used by any of the three circuits ($H$, $X$, $Z$, or $R_Z$) and two-qubit rate $2p$ on $CZ$, reflecting the roughly doubled error rate typical of superconducting two-qubit gates; readout error is not included in the model, so that gate-level depolarizing noise is isolated as the dominant error source. We sweep $p$ over eleven equally spaced values in $[0, 0.05]$, take $3000$ shots per data point, and report the success probability as the empirical frequency with which the 2-bit output register (qubits $q_{17}q_{16}$ in the MBQC and UBQC cases) equals the marked state.

\begin{figure*}[t]
    \centering
    \includegraphics[width=0.72\textwidth]{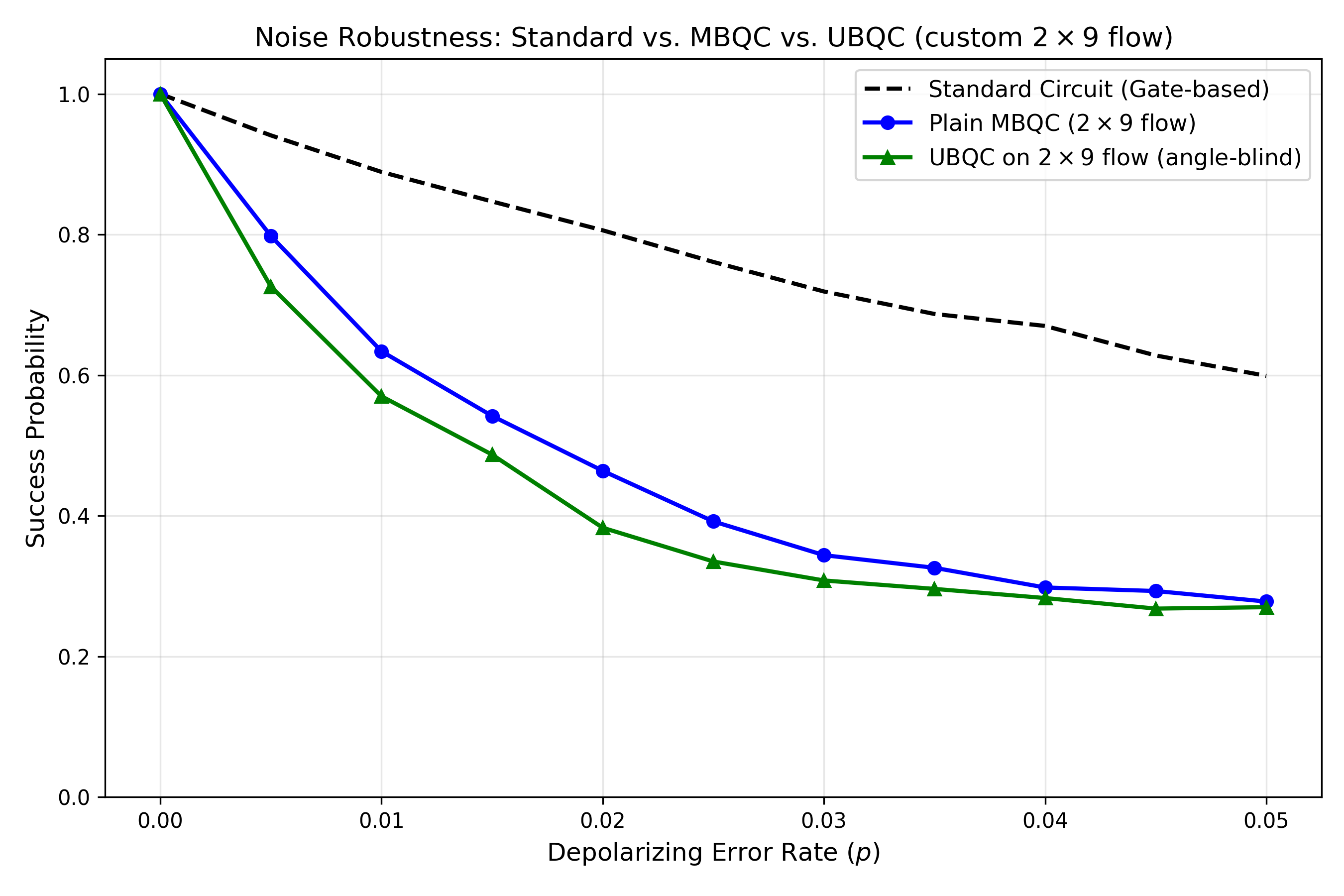}
    \caption{Success probability under depolarizing noise. The standard gate-based circuit (black dashed) remains the most robust, while the UBQC curve (green) stays close to plain MBQC (blue) but is consistently slightly lower, indicating that in the present noise setting the angle-blinding layer contributes only a modest extra gap compared with the dominant MBQC overhead.}
    \label{fig:noise_comparison}
\end{figure*}

Each UBQC data point samples a single fresh realization of the per-vertex pads $(\theta_i, r_i)$ at circuit construction and runs all $3000$ shots on that single realization, so pad-realization variance is conflated with shot noise in the reported curves. This single-realization curve is nonetheless representative: BFK09 correctness (Section~\ref{sec:ubqc_security}) makes the noise-free output pad-independent, and since depolarizing noise is angle-invariant, the only realization-dependent noise channel is the $\mathrm{Binomial}(n, 1/2)$ count of $r_i=1$--conditional $X$ corrections, whose $O(\sqrt{n})$ fluctuation perturbs the gate budget at the same order as our $1/\sqrt{3000}$ shot-noise resolution. All curves in Fig.~\ref{fig:noise_comparison} correspond to the oracle marking $|00\rangle$; the other three oracle settings yield qualitatively identical curves and are omitted for space.

\noindent \textbf{Impact of resource overhead.} The standard circuit-based implementation (dashed black line) exhibits the highest robustness. This is expected, as it uses only 2 physical qubits and a shallow depth, and accumulates fewer errors than the MBQC implementations. In contrast, the MBQC simulations (blue and green lines) require an 18-qubit entangled state and a deeper sequence of measurements, making them more sensitive to noise. This gap quantifies the per-shot error budget one pays for moving a two-qubit algorithm from a two-qubit circuit to an 18-qubit measurement pattern.

\medskip

\noindent \textbf{Blindness with modest extra loss.} The performance of the UBQC protocol (green triangles) remains close to that of the plain MBQC simulation (blue circles), but it is consistently slightly lower across most of the sweep. Within the depolarizing-noise model and single-pad-realization setting used here, this suggests that the additional blinding operations---the per-qubit $R_Z(\theta_i)$ pads applied at preparation, the compensating $R_Z(-\theta_i)$ absorbed into the blinded measurement angle $\delta_i$, and the in-circuit $X$-correction and re-measurement that unblinds qubits with $r_i = 1$---introduce only a modest extra loss relative to plain MBQC. The dominant degradation is still driven by the size and entanglement depth of the underlying MBQC graph, not by the privacy-preserving layer itself.

\subsection{Resource Overhead and Scalability}
\label{sec:resource_overhead}

Section~\ref{sec:robustness} attributed the noise-robustness gap to the size and entanglement depth of the MBQC graph; this section makes that structure explicit.

\noindent \textbf{Scaling formula and 2-qubit Grover footprint.} Our simulation of a 2-qubit algorithm uses 18 physical qubits, and we now derive this count from the MBQC scaling formula. In the MBQC framework, the computation consumes the resource state as it progresses. Unlike the circuit model, where the number of qubits equals the logical register width ($n$), ancillae aside, simulating MBQC on a circuit-based platform requires representing the entire graph state simultaneously. Therefore, the number of qubits required for the simulation, $N_{MBQC}$, scales with the spacetime volume of the computation~\cite{Raussendorf2001, Danos2005}, proportional to the product of the logical circuit width ($n$) and depth ($d$):
\begin{equation}
    N_{MBQC} = O(n\,d).
\end{equation}
For the $2\times 9$ flow used in this paper, this bound is tight: $N_{MBQC}=nd=18$.

For the implemented 2-qubit Grover's algorithm, the logical depth includes initialization, the oracle, and the diffusion operator. Each logical gate operation in MBQC consumes a chain of physical qubits whose size is fixed by its measurement pattern: for the universal gate set of Section~\ref{sec:universality} (Table~\ref{tab:pattern_summary}), $H$ occupies a 2-vertex chain and $R_Z(\theta)$ via the $J(\phi)$ decomposition occupies a 3-vertex chain, while the two-qubit $CZ$ occupies a $2\times 3$ block of $6$ vertices as a standalone pattern. Consequently, the 18 qubits used in our work are exactly the vertices of the $2\times 9$ flow, and agree with the scaling formula for $n=2$ and $d=9$, where $d=9$ counts the horizontal columns of the flow and absorbs the Grover logical depth together with the measurement-pattern overhead.

\medskip

\noindent \textbf{Brickwork alternative and blindness scope.} An alternative layout---most notably the brickwork state $\mc{G}_{n,m}$ employed by the original UBQC construction~\cite{Broadbent2009}, whose column count $m$ plays the role of the depth parameter in the brickwork vertex count and grows proportionally to the two-qubit-gate depth---would yield a comparable $O(nm)$ resource for the same logical circuit; we adopt the denser $2\times 9$ flow here because it was sufficient for the two-qubit Grover algorithm and avoids the identity-pattern padding that the brickwork layout would impose for this specific algorithm.\footnote{Concretely, the two-qubit Grover circuit decomposes into $N_{2q}=2$ two-qubit-gate layers (oracle and diffusion), each of which would occupy one $2\times 4$ brick in the BFK09 construction~\cite{Broadbent2009}. Restricting the layout to those two bricks plus one input column would occupy $2\times(4N_{2q}+1)=18$ qubits, numerically matching our custom flow but still algorithm-specific and therefore not yielding structural blindness. A BFK09 brickwork state $\mc{G}_{n,m}$ that preserves the algorithm-independent brick pattern pads with $k\ge 1$ identity bricks, giving $m=4(N_{2q}+k)+1$ and hiding $N_{2q}$ within a client class of depth $\le N_{2q}+k$; for $N_{2q}=2$ this is $26$ qubits ($k=1$, $m=13$) or $34$ qubits ($k=2$, $m=17$), at or beyond the $\sim$30-qubit ceiling of a typical AerSimulator statevector run used in this work.}

Beyond its resource footprint, the flow choice also determines the blindness scope: a full algorithmic brickwork resource is algorithm-independent and therefore provides \emph{structural blindness}---hiding not just the measurement angles and outcomes but also the algorithm identity by revealing only the universal dimensions $(n, m)$ (Section~\ref{sec:ubqc})---in addition to the angle-based blindness established here (Sections~\ref{sec:ubqc_security} and~\ref{sec:conclusion}). The $G_{2,5}$ experiment of Section~\ref{sec:brickwork_fragment} should be read as a primitive-level brickwork compatibility check: it verifies Clifford, entangling, and non-Clifford building blocks under UBQC-style pads, while leaving full structural blindness to a larger brickwork embedding.

The classical resources grow in parallel, with one measurement outcome bit per graph vertex and a feed-forward depth bounded by the measurement schedule. While this overhead is significant compared to the native circuit-model implementation of the same algorithm, it is an inherent cost of simulating a measurement-based architecture on a gate-based simulator. This approach nonetheless enables verification of MBQC-specific protocols, such as UBQC, that are not natively exposed on standard gate-based cloud processors; Section~\ref{sec:comparison} situates this capability relative to existing MBQC simulation tools.

\subsection{Comparison with Existing Methods}
\label{sec:comparison}

Existing MBQC simulation tools fall into two broad categories. Specialized pattern compilers such as \texttt{graphix}~\cite{graphix}, \texttt{MentPy}~\cite{mentpy}, and the MBQC module of \texttt{Paddle-Quantum}~\cite{paddlemqbc} offer strong graph and flow analysis, but they handle outcome-conditional Pauli byproducts either as classical post-processing once the pattern has been evaluated or as host-side feed-forward that rewrites subsequent measurement angles between shots. Gate-level MBQC demonstrations on general-purpose frameworks, such as the Qiskit-based study of Kashif and Al-Kuwari~\cite{Kashif2022}, typically stop at cluster-state preparation and static measurement patterns without fully integrating the flow-induced corrections inside the circuit. In all of these cases, the correction logic lives \emph{outside} the quantum program.

\noindent \textbf{Why in-circuit correction is the enabling primitive for UBQC.} Our comparison criterion is not merely whether a simulator reproduces the final output distribution of an ideal UBQC run---a post-processing simulator can already do that. Rather, we ask whether a single executable quantum program can follow the measurement-by-measurement BFK09-style update order on the implemented graph. This criterion also keeps the construction close to the primitive set that hardware with mid-circuit measurement and classical feed-forward would need to support, although this paper reports simulator-based validation rather than hardware execution.

Under that criterion, every outcome must be read and acted upon \emph{within the same quantum program}. UBQC makes this unavoidable: at step $i$ the server measures with the blinded angle $\delta_i = \phi'_i + \theta_i + \pi r_i$, returns $s_i$, and $\delta_{i+1}$ is determined by the flow dependency on earlier $s_j$ together with the client's secret pads $(\theta_i, r_i)$. A simulator that commutes the byproduct corrections past the measurements and resolves them as a post-hoc classical layer can still reproduce the MBQC output, but it does not preserve the same single-program execution trace as the client/server update order modeled here.

We emphasise that this requirement is not met merely by invoking Qiskit's \texttt{c\_if}/\texttt{if\_test} construct on isolated gates. A general dynamic-circuit program admits arbitrary runtime classical arithmetic on the stored bits, whereas the flow-based correction formula of Eq.~\eqref{phi2} involves XOR-parity sums over unbounded neighbourhood sets $S_X(i)$ and $S_Z(i)$ that would, if taken literally, require accumulating a classical register and feeding the parity back into the measurement basis. Our construction (Algorithm~\ref{alg:Measurement}) is precisely the observation that this arithmetic can be \emph{decomposed}: each summand $s_j$ contributes an independent single-qubit conditional Pauli on the receiver qubit whose basis or readout is being corrected, and their composition realises the XOR-parity up to a global phase that the subsequent $Z$-basis measurement discards. The MBQC correction engine of Section~\ref{sec:engine} is therefore not a direct translation of Eq.~\eqref{phi2}, but a factorisation of it into an in-circuit subroutine that uses only the minimal \texttt{c\_if} interface and requires no classical-register arithmetic beyond bit-valued conditionals. The per-gate verification experiments of Appendix~\ref{app:gate_verification} confirm that this factorisation is consistent with the textbook Pauli-frame derivations. The UBQC angle-update rule of Eq.~\eqref{eq:blind_angle} then composes with the same factorisation by absorbing the client's secret pads $(\theta_i, r_i)$ into the measurement basis (Algorithm~\ref{alg:BlindMeasurement}) rather than into the correction logic, so that the resulting circuit implements the same update order as the BFK09 exchange on the fixed $2\times 9$ flow of Section~\ref{sec:grover}. In this sense, in-circuit correction is not a stylistic choice but the minimum structural requirement for the dynamic-circuit UBQC simulation studied here.

Under this criterion, simulators that treat corrections as classical post-processing remain valuable for pattern verification, but they target a different execution level from the fixed-graph, in-circuit blind-protocol simulation studied here. Structural blindness from a full brickwork resource remains beyond the qubit budget used here (Sections~\ref{sec:resource_overhead} and~\ref{sec:conclusion}), although Section~\ref{sec:brickwork_fragment} verifies a brickwork universal-gate fragment under the same implementation machinery. Within the fixed-graph angle-blindness regime, none of the representative references compared in Table~\ref{tab:comparison_methods} simultaneously documents (i) flow-based Pauli corrections executed inside the circuit, (ii) a BFK09-style UBQC update layer on top of the same circuit, (iii) a noise-robustness evaluation for that protocol configuration (Section~\ref{sec:robustness}), and (iv) the MBQC-side resource footprint (Section~\ref{sec:resource_overhead}) in a single framework. Table~\ref{tab:comparison_methods} should therefore be read as an execution-level comparison of the cited references used in this paper; for repository citations, it reflects the states cited in the bibliography (accessed on 2026-04-15), rather than a claim about every later software release.

\begin{table*}[tbp]
\centering
\caption{Axis-by-axis comparison of the representative MBQC simulation references cited in this paper. Entries summarize the capabilities documented in the cited work or repository state used for the citation; for repository citations, the comparison is limited to the states accessed on 2026-04-15. Entries marked ``Not documented'' or ``Not reported'' indicate that we did not find an explicit statement in the cited source. ``In-circuit'' means outcome-conditional Pauli corrections are executed inside the quantum program rather than as post-processing. ``Partial'' in the Flow-based Pauli column means that byproducts are handled only for a restricted gate set.}
\label{tab:comparison_methods}
\renewcommand{\arraystretch}{1.25}
\footnotesize
\setlength{\tabcolsep}{4pt}
\begin{tabular}{@{}>{\raggedright\arraybackslash}p{2.1cm}
                >{\raggedright\arraybackslash}p{1.8cm}
                >{\raggedright\arraybackslash}p{2.2cm}
                >{\raggedright\arraybackslash}p{1.6cm}
                >{\raggedright\arraybackslash}p{1.4cm}
                >{\raggedright\arraybackslash}p{2.1cm}
                >{\raggedright\arraybackslash}p{1.4cm}
                >{\raggedright\arraybackslash}p{1.8cm}@{}}
\hline
Reference & Framework & \makecell[l]{Correction\\mode} & \makecell[l]{Flow-based\\Pauli} & \makecell[l]{UBQC\\layer} & \makecell[l]{Evaluation\\target} & \makecell[l]{Noise\\study} & \makecell[l]{BFK09-style\\trace} \\
\hline
\texttt{graphix}~\cite{graphix} & \makecell[l]{Python,\\multi-backend} & \makecell[l]{Classical\\post-processing} & \makecell[l]{Yes\\(auto flow)} & \makecell[l]{Not\\documented} & \makecell[l]{General-purpose,\\pattern-level} & Yes & \makecell[l]{Not\\documented} \\
\texttt{MentPy}~\cite{mentpy} & Python & \makecell[l]{Classical\\post-processing} & Yes & \makecell[l]{Not\\documented} & \makecell[l]{QML /\\parametrized\\MBQC} & Limited & \makecell[l]{Not\\documented} \\
\makecell[l]{Paddle-Quantum\\MBQC~\cite{paddlemqbc}} & PaddlePaddle & \makecell[l]{Not\\documented} & \makecell[l]{Not\\documented} & \makecell[l]{Not\\documented} & \makecell[l]{MBQC module /\\toolkit use cases} & \makecell[l]{Yes\\(toolkit)} & \makecell[l]{Not\\documented} \\
\makecell[l]{Kashif and\\Al-Kuwari~\cite{Kashif2022}} & Qiskit & \makecell[l]{Static\\patterns} & \makecell[l]{Gate-\\specific} & \makecell[l]{Not\\documented} & \makecell[l]{Gate demos,\\teleportation,\\Grover} & \makecell[l]{Not\\reported} & \makecell[l]{Not\\documented} \\
\textbf{This work} & \makecell[l]{Qiskit\\dynamic\\circuits} & \makecell[l]{\textbf{In-circuit}\\(\texttt{c\_if}/\\\texttt{if\_test})} & \makecell[l]{Yes\\($2\times 9$ flow)} & \makecell[l]{\textbf{Yes}\\(angle-\\blind)} & \makecell[l]{2-qubit\\Grover} & \makecell[l]{Yes\\(depol.)} & \textbf{Yes} \\
\hline
\end{tabular}
\end{table*}

\section{Conclusion}
\label{sec:conclusion}
This work shows that \textit{in-circuit} correction is the key structural ingredient in the dynamic-circuit UBQC simulation studied here. By placing both the flow-based Pauli corrections and the UBQC angle-update rule inside the same executable program, the construction follows the BFK09-style client/server update order on the implemented graph. Realized in Qiskit on a custom $2\times 9$ flow, the resulting pipeline reproduces two-qubit Grover deterministically under both plain MBQC and the BFK09 UBQC layer, with fixed-graph angle-based blindness and only a modest additional gap from the blinding layer under the depolarizing-noise setting considered here (Section~\ref{sec:robustness}). The additional $G_{2,5}$ brickwork-fragment experiment shows that the same machinery also supports representative universal-gate primitives, including a $CZ$-equivalent entangling primitive up to a known local Clifford frame.

The algorithm-specific $2\times 9$ flow is a deliberate scope choice: it isolates the fixed-graph angle-blindness mechanism within the qubit budget of current statevector simulators, while structural blindness awaits a full brickwork resource beyond that ceiling (Section~\ref{sec:resource_overhead}). The $G_{2,5}$ result narrows this gap at the primitive level, but it does not claim full algorithmic structural blindness. Even within this restricted scope, the contribution is not confined to post-processing simulation. Because every correction and unblinding step is represented by dynamic-circuit primitives, the program exposes the operations that hardware with mid-circuit measurement and classical feed-forward would need to support, providing a concrete bridge from the BFK09 protocol specification toward hardware-oriented studies of fixed-graph, privacy-preserving delegated quantum computing.

Natural extensions include instantiating UBQC on a full brickwork resource to close the structural-blindness gap, scaling the universal gate library of Table~\ref{tab:pattern_summary} to deeper benchmarks, extending the framework to verifiable BQC variants~\cite{Fitzsimons2017, Kashefi2017, drmota2023verifiable}, and replacing the gate-level depolarizing channel with device-characterized noise models. Together these directions would extend the fixed-graph protocol-update bridge established here toward broader hardware-oriented studies of delegated quantum-computing simulations.

\section*{Data and Code Availability Statement}
Data and code are available from the corresponding author upon reasonable request.

\appendices
\section{Pauli-Frame Derivations for the Universal Gate Set}
\label{app:pauli_frame}

This appendix supplies the step-by-step Pauli-frame derivations corresponding to the measurement patterns listed in Table~\ref{tab:pattern_summary} and referenced throughout Section~\ref{sec:universality}. The derivations are standard (cf. \cite{Raussendorf2001, Danos2005, Wallden2018}) and are reproduced here only for self-containedness and for readers wishing to verify the byproduct operators column-by-column.

\subsection{Hadamard Gate}
Starting from the entangled state of the $q_0$--$q_1$ chain and measuring $q_0$ along $X(0)$, the two-qubit system collapses to $\ket{s_0}_0 \otimes X_1^{s_0} H_1 \ket{\psi}_1$, where $s_0 \in \{0,1\}$ is the measurement outcome. Applying $X^{s_0}$ as a classically controlled correction to $q_1$ yields $H \ket{\psi}$ on the readout qubit.

\subsection{X Gate}
For the three-qubit chain $q_0$--$q_1$--$q_2$ with angles $(\phi_0, \phi_1) = (0, \pi)$: after measuring $q_0$, the remaining state is $\ket{s_0}_0 \otimes X_1^{s_0} H_1 \ket{\psi}_1$. Measuring $q_1$ along $X(\pi)$ then gives $\ket{s_0}_0 \otimes \ket{s_1}_1 \otimes X_2^{s_1} H_2 Z_2 X_2^{s_0} H_2 \ket{\psi}_2$. Using $HZ=XH$ and $XZ=-ZX$, this simplifies to
\begin{equation*}
\ket{s_0}_0 \otimes \ket{s_1}_1 \otimes (-1)\, X_2^{s_1} Z_2^{s_0}\, X_2\, \ket{\psi}_2.
\end{equation*}
Applying the corrections $X_2^{s_1}$ and $Z_2^{s_0}$ implements an $X$-gate on the input up to a global phase.

\subsection{\texorpdfstring{$R_Z(\theta)$-Gate (Z and T as Special Cases)}{RZ(theta)-Gate (Z and T as Special Cases)}}
For the chain $q_0$--$q_1$--$q_2$ with angles $(\phi_0, \phi_1) = (-\theta, 0)$: measuring $q_0$ along $X(-\theta)$ gives $\ket{s_0}_0 \otimes X_1^{s_0} H_1 R_Z(\theta)_1 \ket{\psi}_1$. Measuring $q_1$ along $X(0)$ then yields $\ket{s_0}_0 \otimes \ket{s_1}_1 \otimes X_2^{s_1} H_2 X_2^{s_0} H_2 R_Z(\theta)_2 \ket{\psi}_2$. Using $HX=ZH$, the byproduct operators reduce to $X_2^{s_1} Z_2^{s_0}$ and the readout carries $R_Z(\theta) \ket{\psi}$. The cases $\theta = \pi$ and $\theta = \pi/4$ give the $Z$-gate and $T$-gate, respectively.

\subsection{CZ Gate}
On the $2\times 3$ grid with inputs on $q_0, q_1$ and readouts on $q_4, q_5$: measuring the first row ($q_0, q_2$ along $X(0)$) yields $X_4^{s_2} Z_4^{s_0} \ket{\psi}_4$; measuring the second row ($q_1, q_3$ along $X(0)$) yields $X_5^{s_3} Z_5^{s_1} \ket{\phi}_5$. Accounting for the $CZ$ edge between $q_4$ and $q_5$, the joint output state is
\begin{equation*}
CZ \bigl( X_4^{s_2} Z_4^{s_0} \ket{\psi}_4 \otimes X_5^{s_3} Z_5^{s_1} \ket{\phi}_5 \bigr).
\end{equation*}
Using $CZ(X\otimes I) = (X\otimes Z)CZ$ and $CZ(I\otimes X) = (Z\otimes X)CZ$, together with the fact that $CZ$ commutes with $Z\otimes I$ and $I\otimes Z$, the byproducts can be commuted past $CZ$ to give
\begin{equation*}
\bigl( X_4^{s_2} Z_4^{s_0 \oplus s_3} \otimes X_5^{s_3} Z_5^{s_1 \oplus s_2} \bigr)\, CZ(\ket{\psi}_4 \otimes \ket{\phi}_5),
\end{equation*}
which agrees with the byproduct columns of Table~\ref{tab:pattern_summary}.

\section{Qiskit Verification Figures for the Universal Gate Set}
\label{app:gate_verification}

\begin{figure*}[tbp]
    \centering
    \begin{subfigure}[b]{0.15\textwidth}
        \centering
        \includegraphics[width=\textwidth]{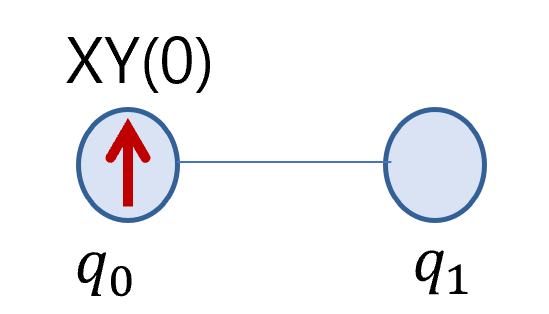}
        \caption{}
        \label{fig:H_gate_sub1}
    \end{subfigure}
    \hfill
    \begin{subfigure}[b]{0.45\textwidth}
        \centering
        \includegraphics[width=\textwidth]{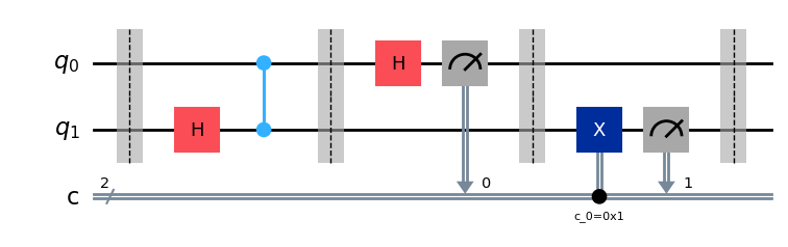}
        \caption{}
        \label{fig:H_gate_sub2}
    \end{subfigure}
    \begin{subfigure}[b]{0.21\textwidth}
        \centering
        \includegraphics[width=\textwidth]{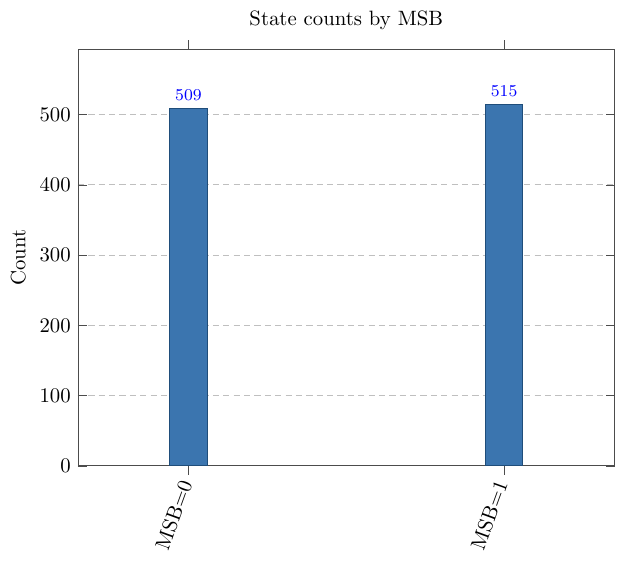}
        \caption{}
        \label{fig:H_gate_sub3}
    \end{subfigure}
    \caption{MBQC simulation using Qiskit: $H$-gate. (a) The $H$-gate pattern, where $q_0$ is the input and $q_1$ is the output; the arrow indicates the measurement angle in the $XY$-plane. (b) Test scenario with $q_0$ initialized in $\ket{0}$. (c) After applying the $H$-gate, $q_1$ collapses to $\ket{0}$ or $\ket{1}$ with equal probability under $Z$-basis readout (empirical: $509/1024 \approx 0.497$ and $515/1024 \approx 0.503$; binomial $1\sigma \approx 0.016$).}
    \label{fig:H_gate}
\end{figure*}

\begin{figure*}[tbp]
    \centering
    \begin{subfigure}[b]{0.21\textwidth}
        \centering
        \includegraphics[width=\textwidth]{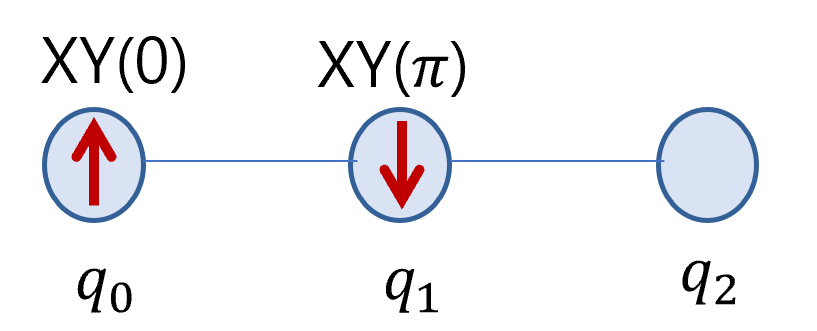}
        \caption{}
        \label{fig:X_gate_sub1}
    \end{subfigure}
    \hfill
    \begin{subfigure}[b]{0.45\textwidth}
        \centering
        \includegraphics[width=\textwidth]{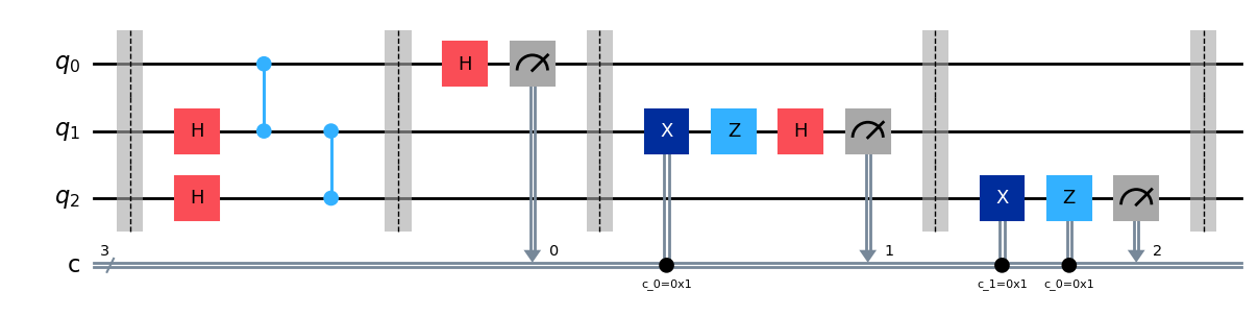}
        \caption{}
        \label{fig:X_gate_sub2}
    \end{subfigure}
    \begin{subfigure}[b]{0.21\textwidth}
        \centering
        \includegraphics[width=\textwidth]{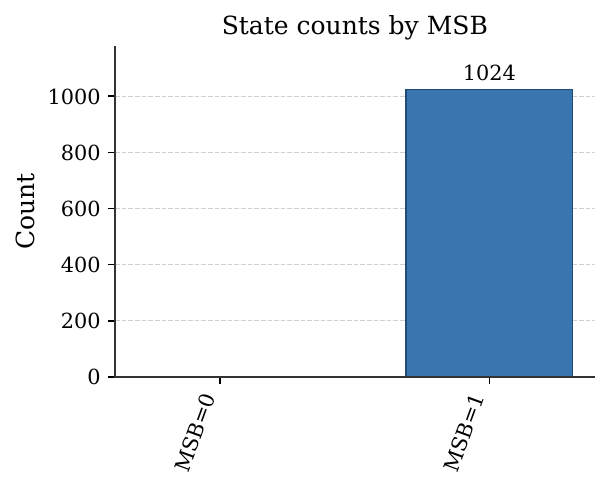}
        \caption{}
        \label{fig:X_gate_sub3}
    \end{subfigure}
    \caption{MBQC simulation using Qiskit: $X$-gate. (a) The $X$-gate pattern, where $q_0$ is the input and $q_2$ is the output; arrows indicate measurement angles in the $XY$-plane. (b) Test scenario with $q_0$ initialized in $\ket{0}$. (c) After applying the $X$-gate, $q_2$ collapses to $\ket{1}$ under $Z$-basis readout (observed $1024/1024$ shots).}
    \label{fig:X_gate}
\end{figure*}

\begin{figure*}[tbp]
    \centering
    \begin{subfigure}[b]{0.21\textwidth}
        \centering
        \includegraphics[width=\textwidth]{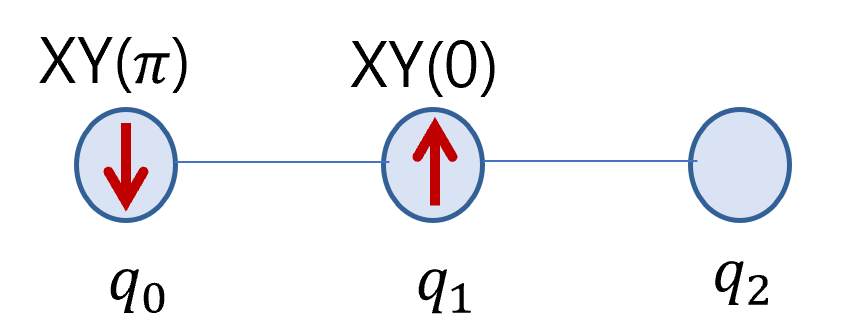}
        \caption{}
        \label{fig:Z_gate_sub1}
    \end{subfigure}
    \hfill
    \begin{subfigure}[b]{0.45\textwidth}
        \centering
        \includegraphics[width=\textwidth]{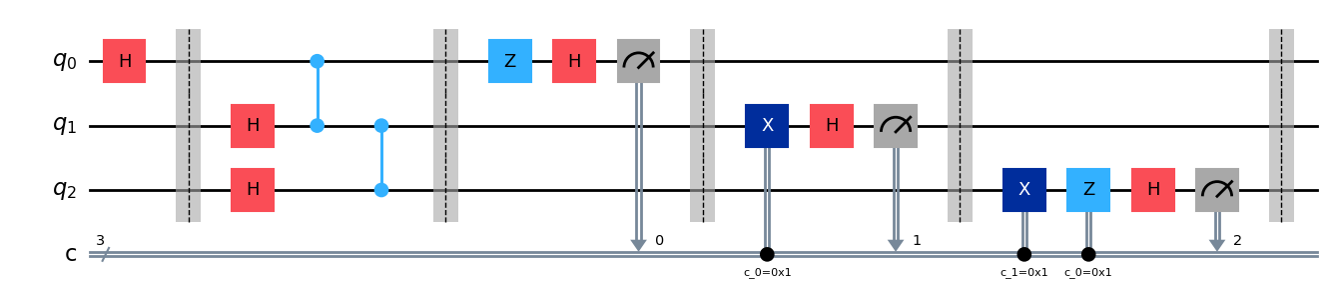}
        \caption{}
        \label{fig:Z_gate_sub2}
    \end{subfigure}
    \begin{subfigure}[b]{0.21\textwidth}
        \centering
        \includegraphics[width=\textwidth]{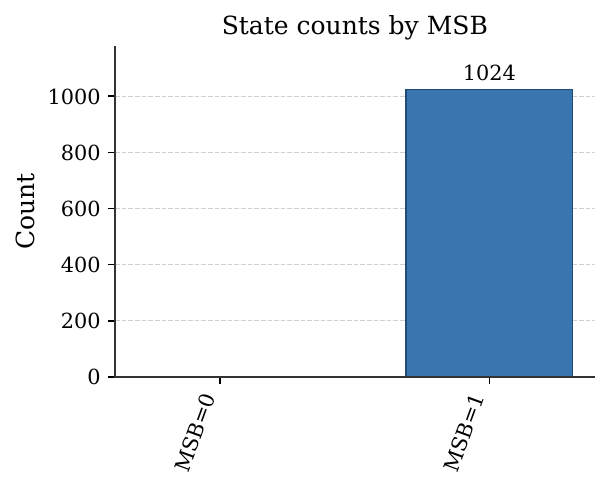}
        \caption{}
        \label{fig:Z_gate_sub3}
    \end{subfigure}
    \caption{MBQC simulation using Qiskit: $Z$-gate. (a) The $Z$-gate pattern, where $q_0$ is the input and $q_2$ is the output; arrows indicate measurement angles in the $XY$-plane. (b) Test scenario with $q_0$ initialized in $\ket{+}$. (c) After applying the $Z$-gate, $q_2$ reaches $\ket{-}$ and collapses to $\ket{1}$ under $X$-basis readout (observed $1024/1024$ shots).}
    \label{fig:Z_gate}
\end{figure*}

\begin{figure*}[tbp]
    \centering
    \begin{subfigure}[b]{0.21\textwidth}
        \centering
        \includegraphics[width=\textwidth]{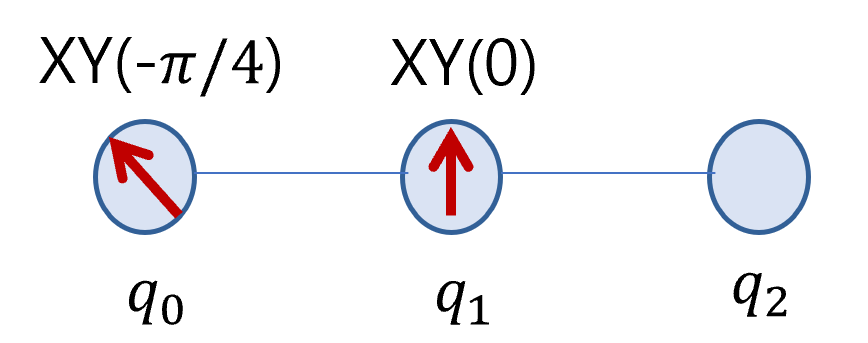}
        \caption{}
        \label{fig:T_gate_sub1}
    \end{subfigure}
    \hfill
    \begin{subfigure}[b]{0.45\textwidth}
        \centering
        \includegraphics[width=\textwidth]{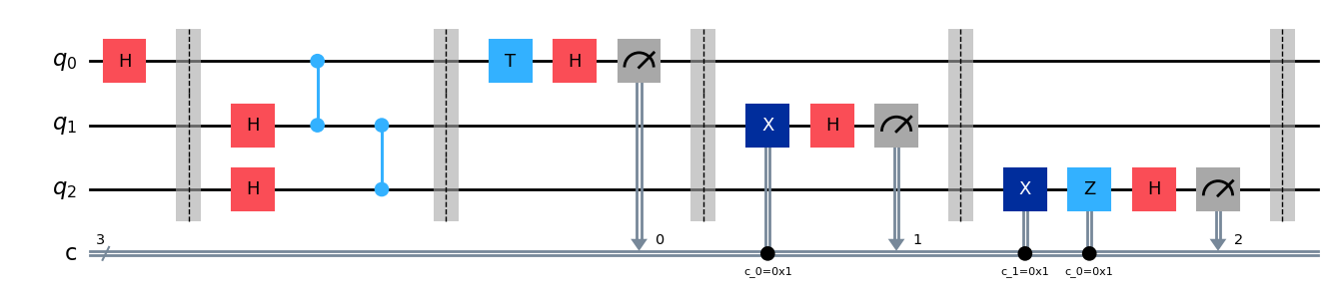}
        \caption{}
        \label{fig:T_gate_sub2}
    \end{subfigure}
    \begin{subfigure}[b]{0.21\textwidth}
        \centering
        \includegraphics[width=\textwidth]{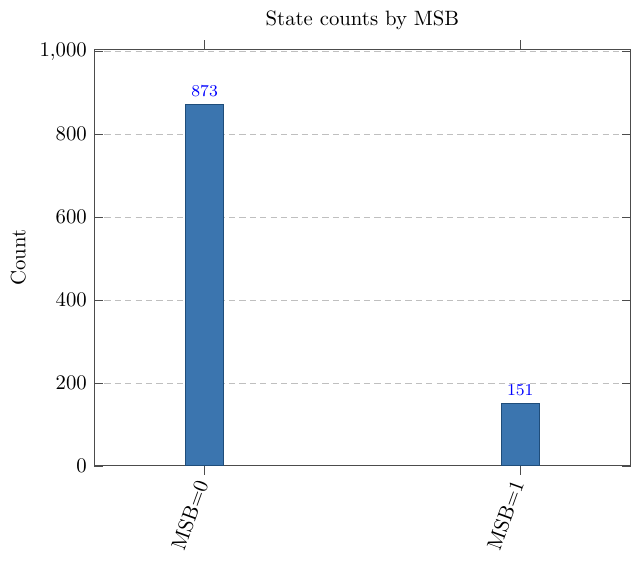}
        \caption{}
        \label{fig:T_gate_sub3}
    \end{subfigure}
    \caption{MBQC simulation using Qiskit: $T$-gate. (a) The $T$-gate pattern, where $q_0$ is the input and $q_2$ is the output; arrows indicate measurement angles in the $XY$-plane. (b) Test scenario with $q_0$ initialized in $\ket{+}$. (c) After applying the $T$-gate, $q_2$ reaches $\ket{+_{\pi/4}} = \tfrac{1}{\sqrt{2}}(\ket{0} + e^{i\pi/4}\ket{1})$ and collapses to $\ket{0}$ with theoretical probability $\cos^2(\pi/8) \approx 0.8536$ under $X$-basis readout (empirical: $873/1024 \approx 0.853$).}
    \label{fig:T_gate}
\end{figure*}

\begin{figure*}[tbp]
    \centering
    \begin{subfigure}[b]{\textwidth}
        \centering
        \centerline{\includegraphics[width=\textwidth]{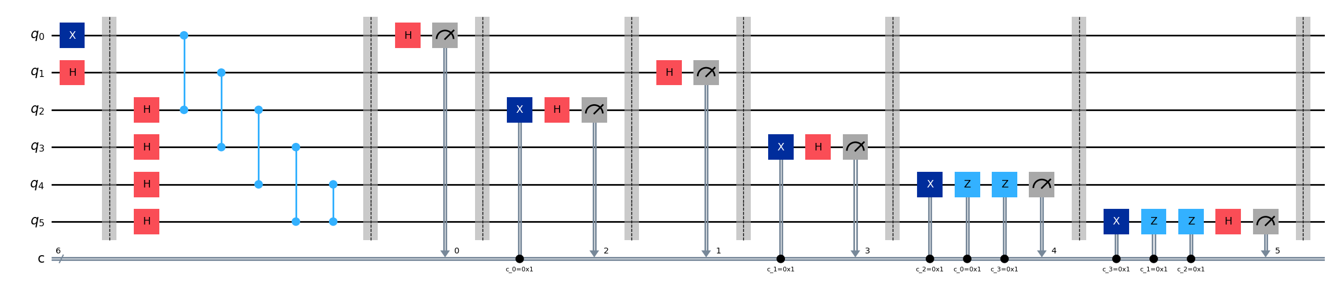}}
        \caption{}
        \label{fig:CZ_gate_sub1}
    \end{subfigure}

    \vspace{0.5cm}

    \begin{subfigure}[b]{0.21\textwidth}
        \centering
        \includegraphics[width=\textwidth]{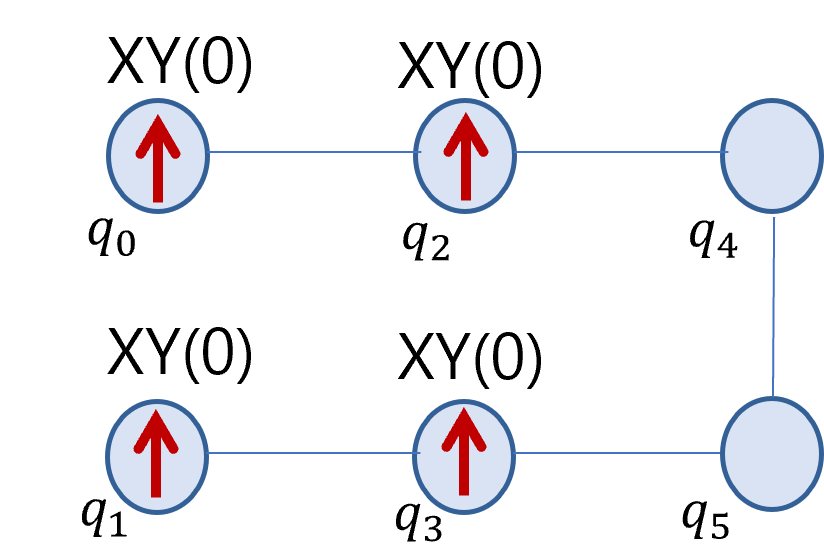}
        \caption{}
        \label{fig:CZ_gate_sub2}
    \end{subfigure}
    \hfill
    \begin{subfigure}[b]{0.21\textwidth}
        \centering
        \includegraphics[width=\textwidth]{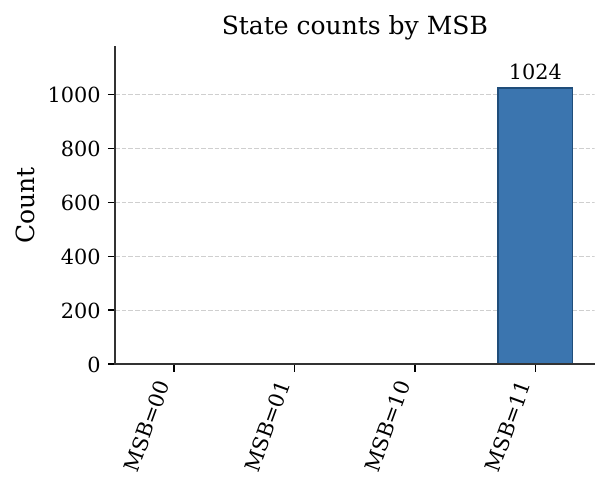}
        \caption{}
        \label{fig:CZ_gate_sub3}
    \end{subfigure}
    \caption{MBQC simulation using Qiskit: $CZ$-gate. (a) The $CZ$-gate pattern, where $q_0$ and $q_1$ are inputs and $q_4$ and $q_5$ are outputs; arrows indicate measurement angles in the $XY$-plane. (b) Test scenario with input $q_0, q_1 = \ket{1}\otimes\ket{+}$. (c) After applying the $CZ$-gate the output state is $\ket{1}\otimes\ket{-}$; both output qubits collapse to $\ket{1}$ under $Z$-basis readout on $q_4$ and $X$-basis readout on $q_5$ (observed $1024/1024$ shots).}
    \label{fig:CZ_gate}
\end{figure*}

This appendix collects the Qiskit verification figures referenced in Section~\ref{sec:universality}, one per gate in the universal set $\{H, X, Z, T, CZ\}$: the $H$-gate in Fig.~\ref{fig:H_gate}, the $X$-gate in Fig.~\ref{fig:X_gate}, the $Z$-gate in Fig.~\ref{fig:Z_gate}, the $T$-gate in Fig.~\ref{fig:T_gate}, and the $CZ$-gate in Fig.~\ref{fig:CZ_gate}. In each of Figs.~\ref{fig:H_gate}--\ref{fig:CZ_gate}, panel~(a) visualizes the resource graph with measurement angles indicated as arrows on the $XY$-plane, panel~(b) shows the equivalent dynamic Qiskit circuit with \texttt{c\_if}-based corrections, and panel~(c) shows the 1024-shot outcome distribution. The numerical outcomes cited in Section~\ref{sec:universality} are reproduced in the respective panel~(c) histograms: $509/1024$ vs.\ $515/1024$ for the $H$-gate (Fig.~\ref{fig:H_gate}), a deterministic $\ket{1}$ readout for the $X$-gate (Fig.~\ref{fig:X_gate}), a deterministic $\ket{-}$ output for the $Z$-gate (Fig.~\ref{fig:Z_gate}), the $\cos^2(\pi/8)\approx 0.853$ bias for the $T$-gate (Fig.~\ref{fig:T_gate}), and a deterministic $\ket{11}$ readout for the $CZ$-gate (Fig.~\ref{fig:CZ_gate}).

\section{Qiskit Dynamic-Circuit Realization of the Two-Qubit MBQC Grover Pattern}
\label{app:qiskit_grover}

This appendix reproduces, in landscape orientation for legibility, the full $18$-qubit Qiskit dynamic circuit that realizes the MBQC two-qubit Grover pattern of Fig.~\ref{fig:MBQC_Grover}. The circuit is generated by Algorithm~\ref{alg:Main} for oracle~=~``00'' and is the object whose measurement statistics are reported in Fig.~\ref{fig:GroverOutcomes}(a) of Section~\ref{sec:grover_impl}.

\section{Qiskit Dynamic-Circuit Realization of the Blind MBQC Two-Qubit Grover Pattern}
\label{app:blind_qiskit_grover}

This appendix reproduces, in landscape orientation for legibility, the full $18$-qubit Qiskit dynamic circuit that realizes the UBQC-instantiated MBQC two-qubit Grover pattern of Section~\ref{sec:ubqc_grover}, run on the custom $2\times 9$ flow rather than on the brickwork resource of the original UBQC~\cite{Broadbent2009}. The circuit is generated by Algorithm~\ref{alg:BlindMain} for oracle~=~``00'' and is the object whose measurement statistics are reported in Fig.~\ref{fig:GroverOutcomes}(b) of Section~\ref{sec:ubqc_results}.

\begin{sidewaysfigure*}[p]
    \centering
    \includegraphics[width=\textheight]{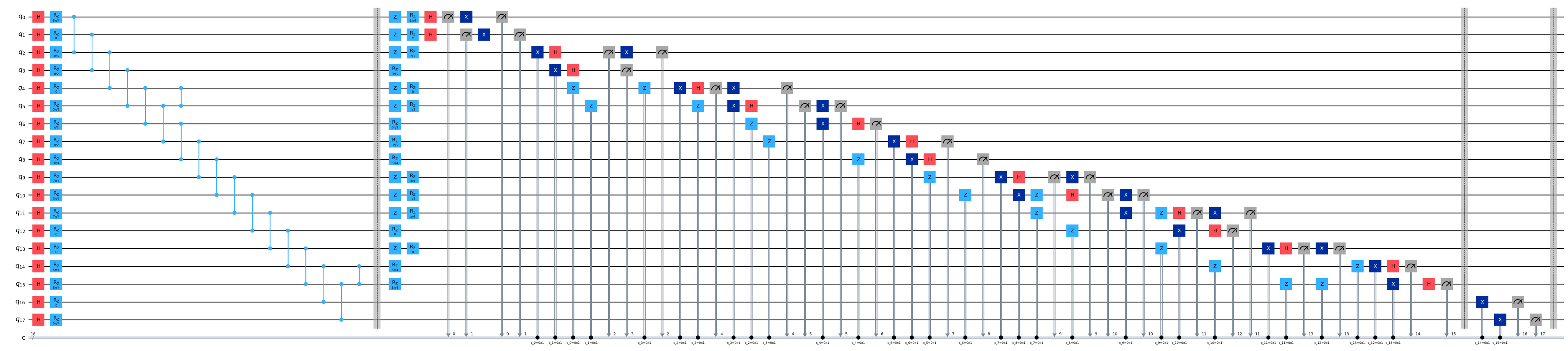}
    \caption{Qiskit dynamic-circuit realization of the blind MBQC two-qubit Grover pattern on the custom $2\times 9$ flow of Section~\ref{sec:grover} (oracle = ``00''); this graph replaces the brickwork resource of the original UBQC~\cite{Broadbent2009} under the simulation-scope caveats of Section~\ref{sec:ubqc}, so angle-based blindness is realized while the graph topology itself remains visible to the server. Each qubit $q_i$ is prepared in the pad-dependent state $\ket{+_{\theta_i}}$ via a Hadamard followed by $R_Z(\theta_i)$ (Algorithm~\ref{alg:BlindInitState}); the $CZ$-entanglement pattern is identical to Fig.~\ref{fig:MBQC2QGrover}. Each measured qubit is preceded by the blinded measurement $M_i^{\delta_i}$ of Eq.~\eqref{eq:blind_delta_measure}, implemented as an optional $Z$ gate (when $r_i=1$), an $R_Z(-\theta_i)$ rotation, and the unblinded measurement primitive $M_i^{\phi'_i}$ of Algorithm~\ref{alg:Measurement}. The post-measurement flip $s'_i\!\leftarrow\! s_i\oplus r_i$ is realized in-circuit by applying an $X$ gate and re-measuring qubit $i$ whenever $r_i=1$, so that downstream \texttt{c\_if} corrections read unblinded outcomes. The output qubits $q_{16}, q_{17}$ receive only the final $X$-correction and are measured directly in the $Z$-basis.}
    \label{fig:BlindMBQC2QGrover}
\end{sidewaysfigure*}

\section{Measurement-Angle Patterns for the \texorpdfstring{$G_{2,5}$}{G2,5} Brickwork Fragment}
\label{app:brickwork_patterns}

This appendix collects the four brickwork-fragment measurement patterns referenced in Section~\ref{sec:brickwork_fragment}. The qubit indexing follows the same flat ordering used in the main-text brickwork discussion: the upper rail is $q_0,q_2,q_4,q_6,q_8$, the lower rail is $q_1,q_3,q_5,q_7,q_9$, and the measured vertices are $q_0,\dots,q_7$.

\begin{figure*}[p]
    \centering
    \includegraphics[width=0.92\textwidth]{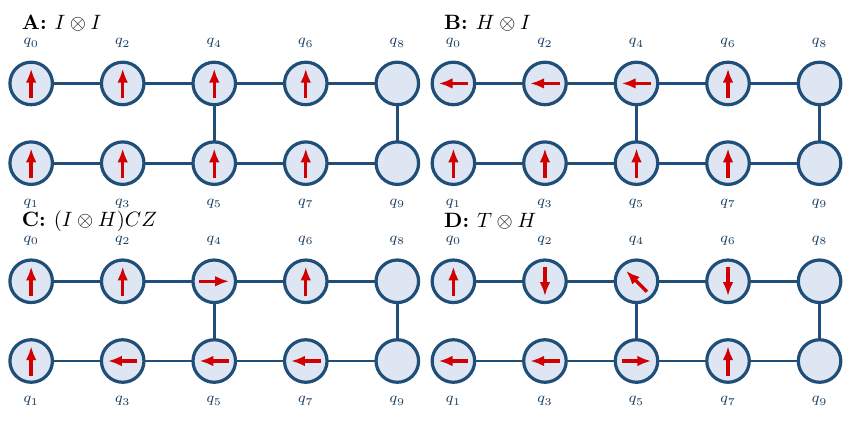}
    \caption{Measurement-angle patterns on the $G_{2,5}$ brickwork-state fragment. Inputs are placed on $q_0,q_1$, outputs on $q_8,q_9$, and the measured vertices are $q_0,\dots,q_7$. Arrows indicate the $XY$-plane measurement angle $\phi=k\pi/4$ used by the adaptive Qiskit measurement routine.}
    \label{fig:brickwork_gate_patterns}
\end{figure*}

\section{Information-Theoretic Bound on Server Leakage}
\label{app:blind_itsec}

This appendix supplies the short derivations of Lemmas~1 and~2 of Section~\ref{sec:ubqc_security}, which together bound what the server can learn about the oracle through the blinded angles and returned outcomes. The argument specialises the standard BFK09 blindness reduction~\cite{Broadbent2009} to the angle alphabet $\mathcal{A} = \{k\pi/4\}_{k=0}^{7}$ and the $2\times 9$ flow of Section~\ref{sec:grover}; the point of reproducing it here is only to make the scope of our blindness claim explicit for the custom-graph setting.

\vspace{0.2cm}

\noindent \textbf{Proof of Lemma~1 (angle pad).}
The client draws $\theta_i \in_R \mathcal{A}$ uniformly and $r_i \in_R \{0,1\}$ uniformly, independently of $\phi'_i$ and of all other pads. Since $\mathcal{A}$ is a cyclic subgroup of $\mathbb{R}/2\pi\mathbb{Z}$ of order $8$ and $\pi$ is the element of order $2$ in it, $\theta_i + \pi r_i \pmod{2\pi}$ is uniform on $\mathcal{A}$: adding $\pi$ to a uniform element of $\mathcal{A}$ yields a permutation of $\mathcal{A}$ (shift by $4$ positions), so the mixture $\tfrac{1}{2}[\theta_i] + \tfrac{1}{2}[\theta_i + \pi]$ is still uniform. Adding the deterministic offset $\phi'_i \in \mathcal{A}$ is a further cyclic shift, so $\delta_i = \phi'_i + \theta_i + \pi r_i \pmod{2\pi}$ is uniform on $\mathcal{A}$ and, a fortiori, statistically independent of $\phi'_i$. Consequently $H(\delta_i \mid \phi'_i) = H(\delta_i) = \log_2 8 = 3$ bits and $I(\phi'_i\,;\,\delta_i) = 0$.

\vspace{0.2cm}

\noindent \textbf{Proof of Lemma~2 (outcome pad).}
Let $s'_i \in \{0,1\}$ be the unblinded outcome that would be produced by measuring qubit $i$ in the basis determined by $\phi'_i$; this is a (possibly non-trivial) function of $\Phi$ and of the prior outcomes $\{s'_j\}_{j<i}$. The server, however, sees $s_i = s'_i \oplus r_i$, and $r_i \in_R \{0,1\}$ is drawn fresh at the client, independent of $\Phi$, $\{s'_j\}_{j<i}$, and all other $r_j$. One-time-pad encryption therefore applies pointwise: for any realisation of $s'_i$, the marginal $P(s_i \mid s'_i) = \tfrac{1}{2}$, so $H(s_i \mid s'_i) = 1$ and $I(s_i\,;\,s'_i) = 0$. Because the $\{r_i\}$ are mutually independent, the server-visible joint distribution $P(\{s_i\})$ is the uniform product distribution on $\{0,1\}^{N_m}$, independent of $\Phi$, where $N_m$ is the number of non-output vertices.

\vspace{0.2cm}

\noindent \textbf{Joint bound.}
Combining the two lemmas, the server-visible transcript $\mathcal{T}_{\mathrm{srv}} = (\{\delta_i\}, \{s_i\}, G, E_{CZ})$ has angle component uniform on $\mathcal{A}^{N_m}$ (Lemma~1) and outcome component uniform on $\{0,1\}^{N_m}$ (Lemma~2), where $N_m=|V\setminus O|$ is the number of measured non-output vertices; both components are independent of the unblinded pattern $\Phi$. Hence
\begin{equation}
    I\!\left(\Phi\,;\,\mathcal{T}_{\mathrm{srv}} \setminus (G, E_{CZ})\right) = 0,
\end{equation}
i.e., conditional on the public graph topology, the server's transcript is distributed identically under all four oracles, and no statistical test---\emph{including} the measurements actually performed in our Qiskit run---can distinguish them. This is the scope of blindness claimed in Section~\ref{sec:ubqc_security}; the complementary channel through $(G, E_{CZ})$ is analysed there as algorithm-structure leakage.

\bibliographystyle{IEEEtran}
\bibliography{references}

\end{document}